\documentclass[aps,prl,twocolumn]{revtex4-1}
\usepackage[latin1]{inputenc}
\usepackage{amsmath}
\usepackage[margin=0.5in]{geometry}
\usepackage{color}
\usepackage{grffile}
\usepackage{hyperref}
\usepackage{float}

\usepackage{amsfonts}
\usepackage{amssymb}
\usepackage{graphicx}
\begin{document}

\newcommand\plotSize{0.485\linewidth}
\title{Identifying nodal properties that are crucial for the dynamical robustness of multi-stable networks}
\author{Pranay Deep Rungta, Chandrakala Meena
and Sudeshna Sinha\thanks{e-mail: sudeshna@iisermohali.ac.in}}

\address{Indian Institute of Science Education and Research
   (IISER) Mohali, SAS Nagar, Sector 81, Mohali 140 306, Punjab,
   India}

\begin{abstract}
        We investigate the collective dynamics of bi-stable elements connected in different network topologies, ranging from rings and small-world networks, to scale-free networks and stars. We estimate the dynamical robustness of such networks by introducing a variant of the concept of multi-node basin stability, which allows us to gauge the global stability of the dynamics of the network in response to local perturbations affecting a certain class of nodes of a system. We show that perturbing nodes with high closeness and betweeness-centrality significantly reduces the capacity of the system to return to the desired state. This effect is very pronounced for a star network which has one hub node with significantly different closeness/betweeness-centrality than all the peripheral nodes. In such a network, perturbation of the single hub node has the capacity to destroy the collective state. On the other hand, even when a majority of the peripheral nodes are strongly perturbed, the hub manages to restore the system to its original state, demonstrating the drastic effect of the centrality of the perturbed node on the dynamics of the network. Further, we explore explore Random Scale-Free Networks of bi-stable dynamical elements. We exploit the difference in the distribution of betweeness centralities, closeness centralities and degrees of the nodes in Random Scale-Free Networks with $m=1$ and $m=2$, to probe which centrality property most inluences the robustness of the collective dynamics in these heterogeneous networks. Significantly, we find clear evidence that the betweeness centrality of the perturbed node is more crucial for dynamical robustness, than closeness centrality or degree of the node. This result is important in deciding which nodes to safeguard in order to maintain the collective state of this network against targetted localized attacks. 
\end{abstract}

\maketitle


Collective spatiotemporal patterns emerging in dynamical networks are determined by the interplay of the dynamics of the nodes and the nature of the interactions among the nodes. So it is of utmost relevance to ascertain what properties of the nodes of the network impact collective dynamics. This also helps to address the important reverse question: perturbation of what class of nodes in the network have the most significant effect on the resilience of the network? Understanding this will allow us determine which nodes render the network most susceptible to external influences. Alternately, it will suggest which nodes to protect more stringently from perturbations in order to protect the dynamical robustness of the entire network. As a test-bed for understanding this we consider the collective dynamics of a group of coupled bi-stable elements. Bi-stable systems are relevant in a variety of fields, ranging from relaxation oscillators\cite{relax_ocs} and multivibrators\cite{multiVib}, to light switches\cite{optical_bistable} and Schmitt triggers. Further it is of utmost importance in digital electronics, where binary data is stored using bi-stable elements. 

Specifically, in this work we will explore bi-stable elements, connected in different network topologies, ranging from regular rings to random scale-free and star networks. We focus on the response of this network to localized perturbations on a sub-set of nodes. The central question we will investigate here is the following: {\em what characteristics of the nodes (if any) significantly affect the global stabilty?} So we will search for discernable patterns amongst the nodes that aid the maintenance of the stability of the collective dynamics of the network on one hand, and the nodes that rapidly destroy it on the other. In particular, we consider three properties of the nodes: degree, betweeness centrality and closeness centrality. Since these features of a node determine the efficiency of information transfer originating from it, or through it, they are expected to influence the propagation of perturbations emanating from the node.

Normalized degree of a node $i$ in an undirected network is given by the number of neighbors that are directly connected to the node scaled by the total number of nodes $N$, and is denoted by $k_i$. So a high degree node indicates that there is direct contact with a larger set of nodes. Normalized betweeness centrality of a node $i$ \cite{chen,pre_hong} is given as: 
$$b_i= \frac{2}{(N-1)(N-2)} \sum_{s,t\in I}\frac{\sigma(s,t|i)}{\sigma(s,t)}$$ 
where $I$ is the set of all nodes, $\sigma(s,t)$ is the number of shortest paths between nodes $s$ and $t$ and $\sigma(s,t|i)$ is the number of shortest paths passing through the node $i$. So if node $i$ has high betweeness-centrality, it implies that it lies on many shortest paths, and thus there is high probability that a communication from $s$ to $t$ will go through it. Normalized Closeness Centrality is defined as:
$$c_i=\frac{N-1}{\sum_{j} d(j,i)}$$
where $d(j,i)$ is the shortest path between node $i$ and node $j$ in the graph. Namely, it is the inverse of the average length of of the shortest path between the node and all other nodes in the network\cite{bavelas}. So high closeness centrality indicates short communication path to other nodes in the network, as there are minimal number of steps to reach other nodes.

In this work we will explore the extent to which the features of the nodes given above influence the recovery of a network from large localized perturbations. In order to  gauge the global stability and robustness of the collective state of this network, we will {\em  introduce a variant of the recent framework of multi-node basin stability} \cite{mitra}. In general, the basin stability of a particular attractor of a multi-stable dynamical system is given by the fraction of perturbed states that return to the basin of the attraction of the dynamical state under consideration. In our variant of this measure, we consider an initial state where all the bi-elements in the network are in the same well, and we will refer to this as a {\em synchronized state}. So a synchronized state here does not imply complete synchronization. Rather it implies a collective state where the states of the nodes are confined to the neigbourhood of the same attracting stable state, i.e. lies within the basin of attraction of one of the two attracting states. We then perturb a specific number of nodes of a prescribed type, with the perturbations chosen randomly from a given subset of the state space. The multi-node basin stability (BS) is then defined as the fraction of such perturbed states that manage to revert back to the original state from these localized perturbations. Namely, multi-node BS reflects the fraction of the volume of the state space of a sub-set of nodes that belong to the basin of attraction of the synchronized state. So the importance of multi-node BS stems from the fact that it determines the probability of the system to remain in the basin of attraction of the synchronized state when random perturbations affect a specific number of nodes. This allows us to extract the contributions of individual nodes to the overall stability of the collective behaviour of the dynamical network. Further, since one perturbs subsets of nodes with certain specified features, our variant of the concept of multi-node BS will suggest which nodal properties make the network more vulnerable to attack.

	



Specifically we consider the system of $N$ diffusively coupled bi-stable elements, whose dynamics is given as:
	
\begin{equation}
\dot{x_i} = F(x_i) + C \frac{1}{K_i} \sum_j (x_j - x_i) = F(x_i) + C (\langle x_i^{nbhd} \rangle - x_i)
\label{main}
\end{equation}
	
where $i$ is the node index ($i = 1, \dots N$) and $C$ is the coupling constant reflecting the strength of coupling. The set of $K_i$ neighbours of node $i$ depends on the topology of the underlying connectivity, and this form of coupling is equivalent to each site evolving diffusively under the influence of a ``local mean field'' generated by the coupling neighbourhood of each site $i$, $\langle x_i^{nbhd} \rangle = \frac{1}{K_i} \sum_j x_j$, where $j$ is the node index of the neighbours of the $i^{th}$ node, with $K_i$ being the total number of neighbours of the node.

The function $F(x)$ gives rise to a double well potential, with two stable states $x^*_-$ and $x^*_+$. For instance one can choose
	$$F(x) = x - x^3$$
	yielding two stable steady states $x^*_{\pm}$ at $+1$ and $-1$, separated by an unstable steady state at $0$. Note that the synchronized state here is a fixed point, either  $x^*_-$ and $x^*_+$, for all the nodes, i.e. $x_i$ is equal to $x^*_-$ or $x^*_+$, for all $i$.
  	
We first investigate the two limiting network cases: (i) {\em Ring}, where all nodes have the same degree, closeness and betweeness centrality, and a (ii) {\em Star network}, where the central (hub) node has the maximum normalized degree ($k_{hub} = 1$), betweeness centrality ($b_{hub} \sim 1$), and closeness centrality ($c_{hub} = 1$), while the rest of the nodes, namely the peripheral nodes (``leaves'') have very low degree ($k_{peri} \sim 0$ for large networks), betweeness centrality ($b_{peri}=0$) and closeness centrality $c_{peri} \sim 0.5$. So on one hand we have the Ring which is completely homogeneous, and on the other hand we have the Star network where the difference in degree, closeness and betweeness centrality of the hub and the peripheral nodes is extremely large. Exploring these limiting cases allows us to gain understanding of the robustness of the network to large perturbations affecting nodes with different properties. 

As indicated eralier, to gauge the effect of different nodal features on the robustness of the dynamical state of the network, we do the following: we first consider a network close to a stable synchronized state, namely one where the states $x_i$ of all the nodes $i$ have a small spread in values centered around  $x^*_-$ or $x^*_+$, i.e. all elements are confined to the same well. We then give a {\em large perturbation} to a small fraction of nodes, denoted by $f$. This strong perturbation typically kicks the state of the perturbed nodes to the basin of attraction of the other well.  We then ascertain whether all the elements return to their original wells after this perturbation, i.e. if the perturbed system recovers completely to the initial state. We repeat this ``experiment'' over a large sample of perturbed nodes and perturbation strengths, and find the fraction of times the system manages to revert to the original state. This measure of global stability is then a variant of multi-node Basin Stability and it is indicative of the robustness of the collective state to perturbations localized at particular nodes of a certain type in the network.\\


{\bf Dynamics of a Ring of Bistable Systems}\\

We first investigate the spatiotemporal evolution of a ring of bistable elements, all of whose states are confined to the same well, other than a few nodes that experience a large perturbation which pushes their state to the basin of the other well. 
We find that even when the fraction of perturbed nodes is very small, these perturbed nodes are unable to return to the original well. That is, the elements in the Ring are unable to drag the few perturbed nodes back to the well of the majority of the elements, suggesting that the Ring is not robust against such localized perturbations. 

\begin{figure}[htb]
	\centering
	\includegraphics[width=0.9\linewidth]{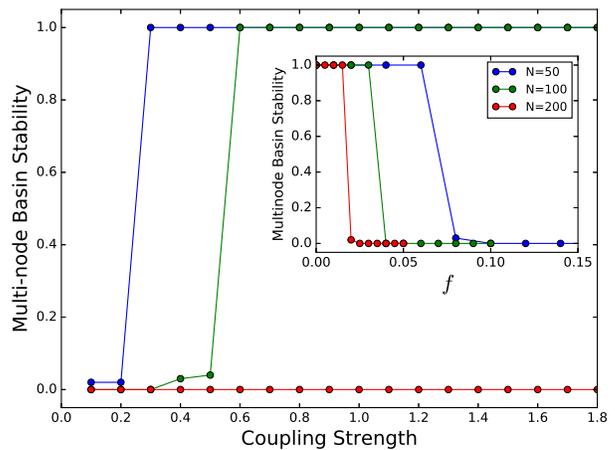}
	
	(a)
	
	\includegraphics[angle=270, width=0.9\linewidth]{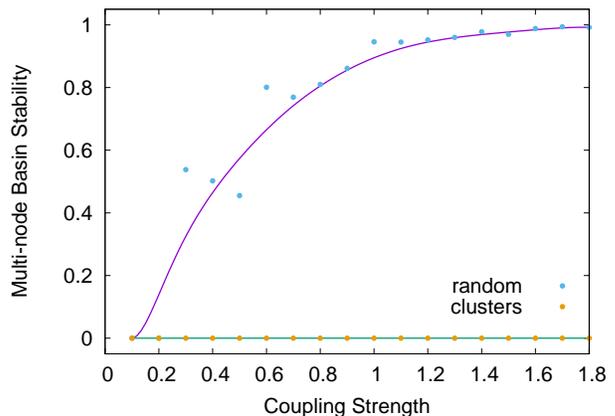}
	
	(b)
	
	\caption{Dependence of Multi-Node Basin Stability on coupling strength, for a ring of bistable elements given by Eqn.~\ref{main}, with the number of perturbed nodes $f$ equal to $0.01$ (blue), $0.02$ (green) and $0.08$ (red). Here the size of the ring is $N=100$, and size of the coupling neighbourhood is $k=2$, namely each site couples to its two nearest neighbours. Panel (a) shows the case where the perturbed nodes occur in clusters, with the inset in (a) showing the dependence of multi-node basin stability on the fraction of perturbed nodes $f$ for different system sizes $N$, for $C=1$. Panel (b) shows the  multi-node basin stabilityfor the case of perturbations on randomly located nodes, for $f=0.08$. The case of perturbation in clusters (orange) is also shown for reference, for the same fraction of perturbed nodes.}
	\label{bsvsc_ring}
\end{figure}

Next we attempt to discern the effect of coupling on the robustness of the dynamics. Fig.~\ref{bsvsc_ring}(a) shows the multi-node basin stability for this system, as the coupling strength is increased  in the range $0$ to $2$, for clusters of perturbed nodes with $f$ ranging from $0.01$ to $0.08$. It is evident from the basin stability of the system, that there is a {\em sharp transition} from zero basin stability, namely the situation where {\em no} perturbed state returns to the original state, to basin stability close to one, namely where {\em all} sampled perturbed states return to the original state. This indicates that the {\em system recovers from large localized perturbations more readily if it is strongly coupled}. Further, the figure also demonstrates the extreme sensitivity of basin stability to the number of nodes being perturbed. We find that the system fails to return to the original state, even at very high coupling strengths, when more than 5\% of the nodes experience perturbations. For instance, Fig.~\ref{bsvsc_ring}(a) shows the case of a single perturbed node (i.e. $f=0.01$), where the entire network recovers for coupling strengths stronger than approximately $0.2$. In contrast, for $f=0.08$, where a cluster of $8$ nodes are perturbed in the Ring of $100$ elements, there is zero basin stability in the entire coupling range. So a Ring loses its ability to return to the original state rapidly with increasing number of perturbed nodes. 

For very small $f$, the entire network returns to its original state, and basin stability is close to $1$. On increasing $f$ one observes that there exists a minimum fraction, which we denote as $f_{crit}$, after which the basin stability sharply declines from $1$. So $f_{crit}$ indicates the minimum fraction of nodes one typically needs to perturb in order to destroy the collective state where all elements are in the same well. We find that high coupling strengths increase $f_{crit}$. For instance, $f_{crit} \approx 0.02$ for $C=0.5$  and $f_{crit} \approx 0.04$ for $C=1$. So for stronger coupling, the bulk of the elements are capable of pulling the perturbed nodes back to original well, increasing the resilience of the network. 

Due to the structure of the ring, the stability of the system with respect to localized perturbations depends on whether the perturbed nodes are contiguous and occur in a cluster (cf. the case in Fig.~\ref{bsvsc_ring}(a)) or randomly spread over the ring, where the locations of the perturbed nodes are uncorrelated. Fig \ref{bsvsc_ring}(b) shows the multi-node basin stability when nodes perturbed are chosen randomly for different values of coupling. We observe that the system is more stable here, as compared to the case when nodes are perturbed in cluster, namely {\em perturbations at random locations in a ring allows the system to recover its original dynamical more readily than perturbations on a cluster of contiguous nodes}.

Further, the inset in Fig. ~\ref{bsvsc_ring}(a) shows the dependence of multi-node basin stability on the fraction of perturbed nodes $f$ for different system sizes $N$. We find that $f_{crit} \sim \frac{C}{N}$, implying that $f_{crit} \rightarrow 0$ as system size $N \rightarrow \infty$. This indicates that in a very large Ring, even the smallest finite fraction of perturbed nodes can disturb the Ring from its original steady state. So one can conclude that the synchronized state in the Ring is very susceptible to destruction, as only very few nodes in the system need to be perturbed in order to push the system out of the original state. Namely, in a ring of bi-stable elements, {\em the collective state where all elements are in the same well, is a very fragile state.}  \\

{\bf Dynamics of a Star Network of Bistable Systems}\\

	Now we study the spatiotemporal evolution of bistable elements connected in a star configuration. Here the central hub node has the maximum degree, betweenness and closeness centrality, while the rest of the nodes, namely the peripheral leaf nodes have very low degree, betweeness and closeness centrality. Namely, in this network the difference in degree, closeness and betweeness centrality of the hub and the peripheral nodes is extremely large. So this network offers a good test-bed to investigate the correlation between specific properties of a node and the resilience of the network to large localized perturbations at such nodes. 

\begin{figure}[htb]
		\centering
		\includegraphics[width=\plotSize]{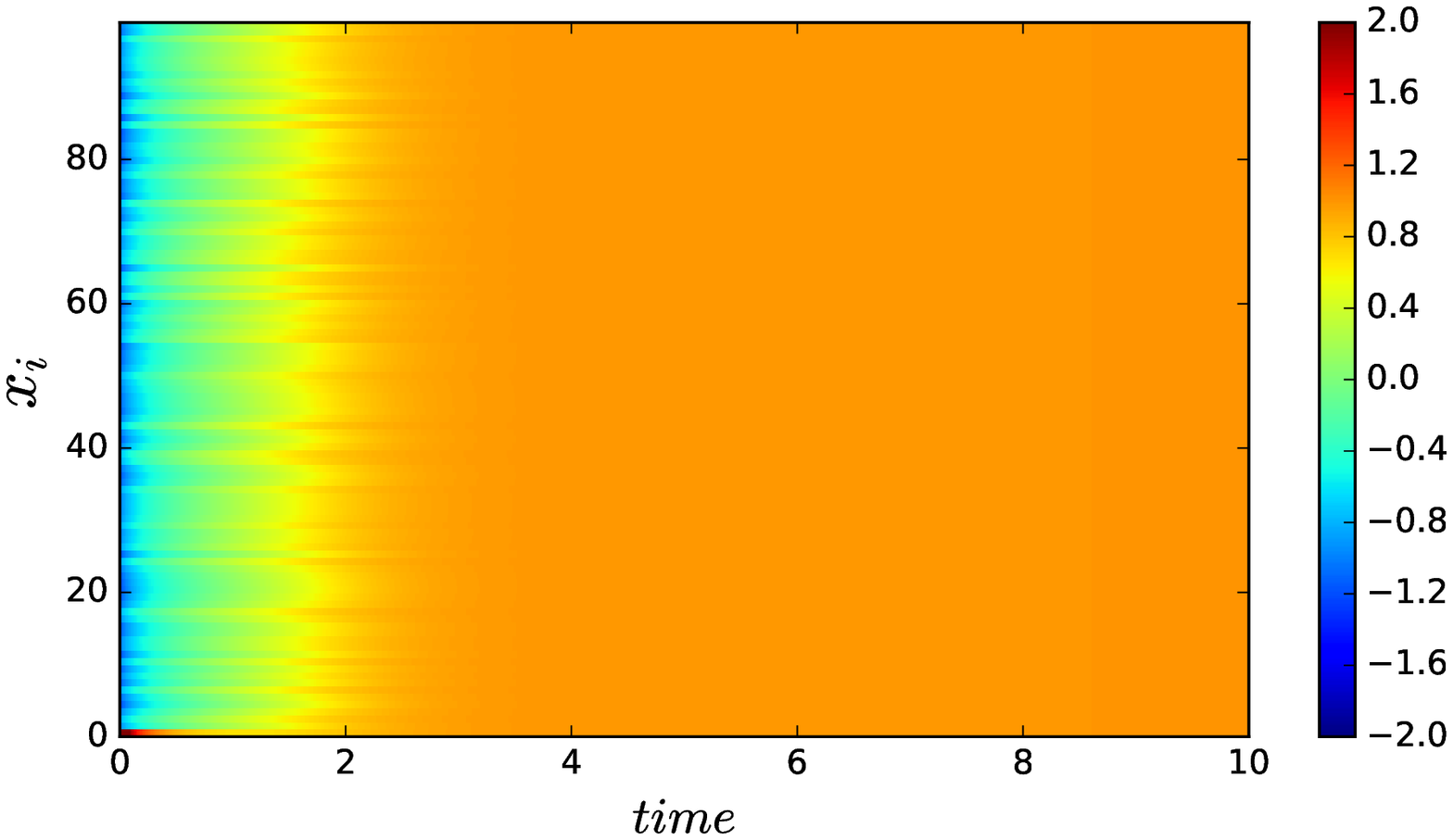}
		\includegraphics[width=\plotSize]{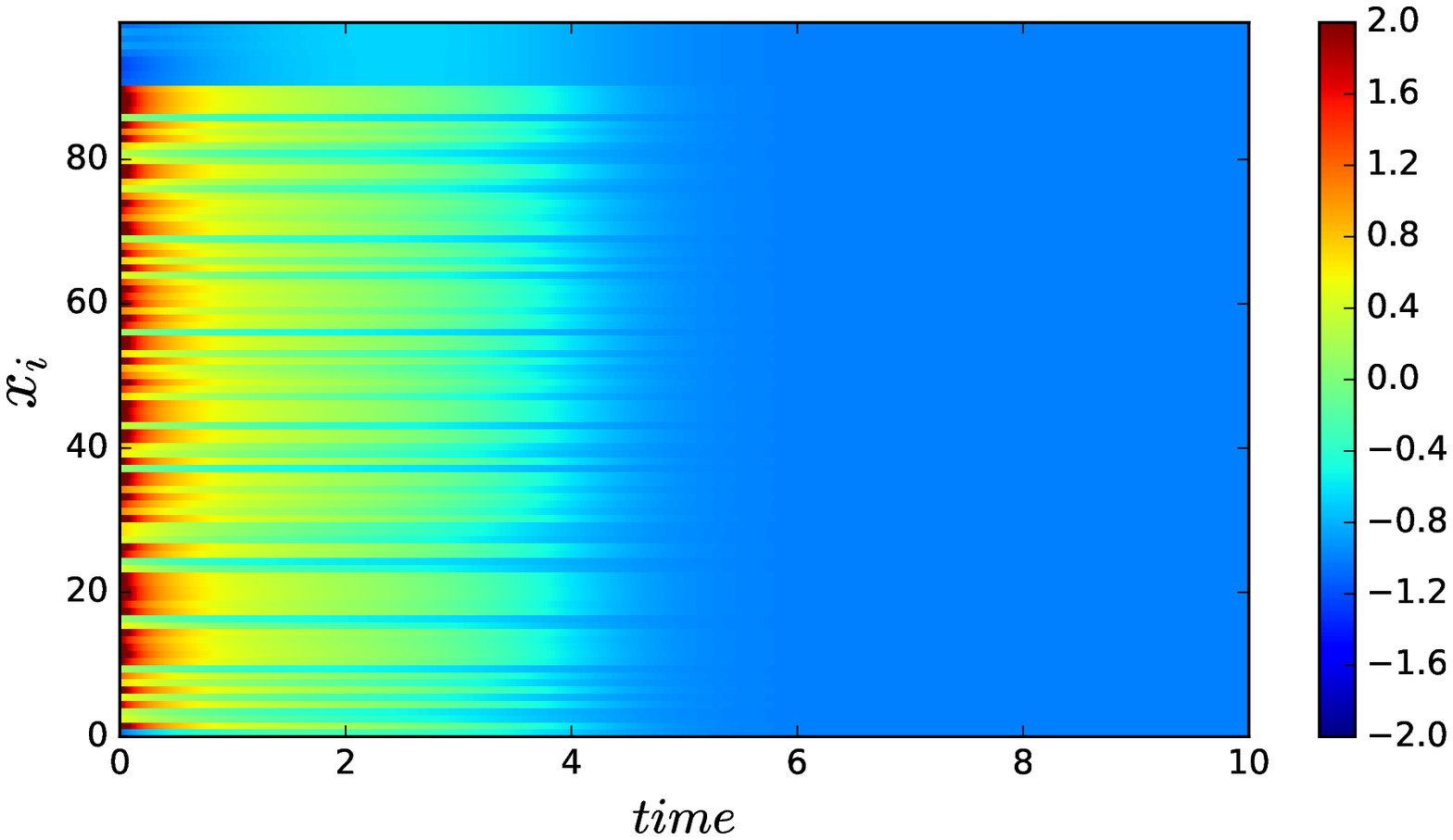}

\hspace{5cm} (a) \hfill (b) \hspace{5cm}
		\caption{Time evolution of $100$ bistable elements coupled in star configuration, given by Eqn.~\ref{main}, with coupling strength $C = 1$. In (a) only the hub node is perturbed; in (b) $90$ peripheral nodes are perturbed.}
		\label{Starspt}
	\end{figure}

Figs.~\ref{Starspt}a-b display the dynamics for two illustrative cases. In Fig.~\ref{Starspt}(a), {\em only the hub node is perturbed} in the star network consisting of $100$ elements. We notice, that this {\em single} perturbed node pulls {\em all} the other nodes of the network away from its original state. So the star network is {\em extremely vulnerable to perturbations at the hub}, and cannot typically recover from disturbances to the state of the hub, even if all the other nodes are unperturbed. On the other hand, Fig.~\ref{Starspt}(b), shows what ensues when a {\em large number of peripheral nodes are perturbed}. Now, even when as many as $90$ nodes are perturbed, namely 90\% of the network experiences a disturbance in its state, the entire network still manages to recover to its original state. This dramatic difference in the outcome of perturbations clearly illustrates how sensitively the robustness of a dynamical state depends on the degree, closeness and betweeness centrality of the perturbed node.

\begin{figure}[htb]
		\centering
		\includegraphics[angle=270, width=0.9\linewidth]{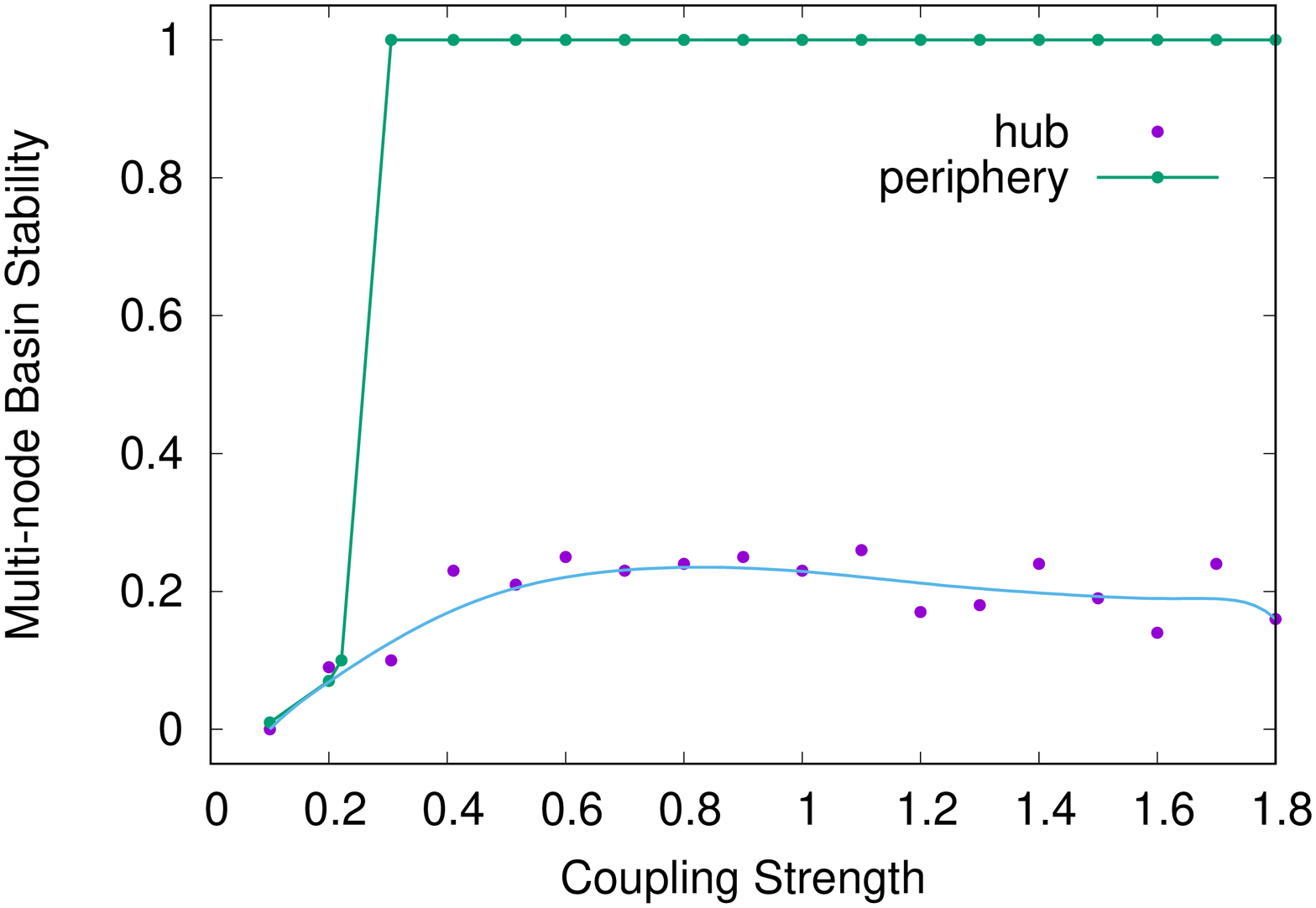}\\(a)\\
		\includegraphics[width=0.9\linewidth]{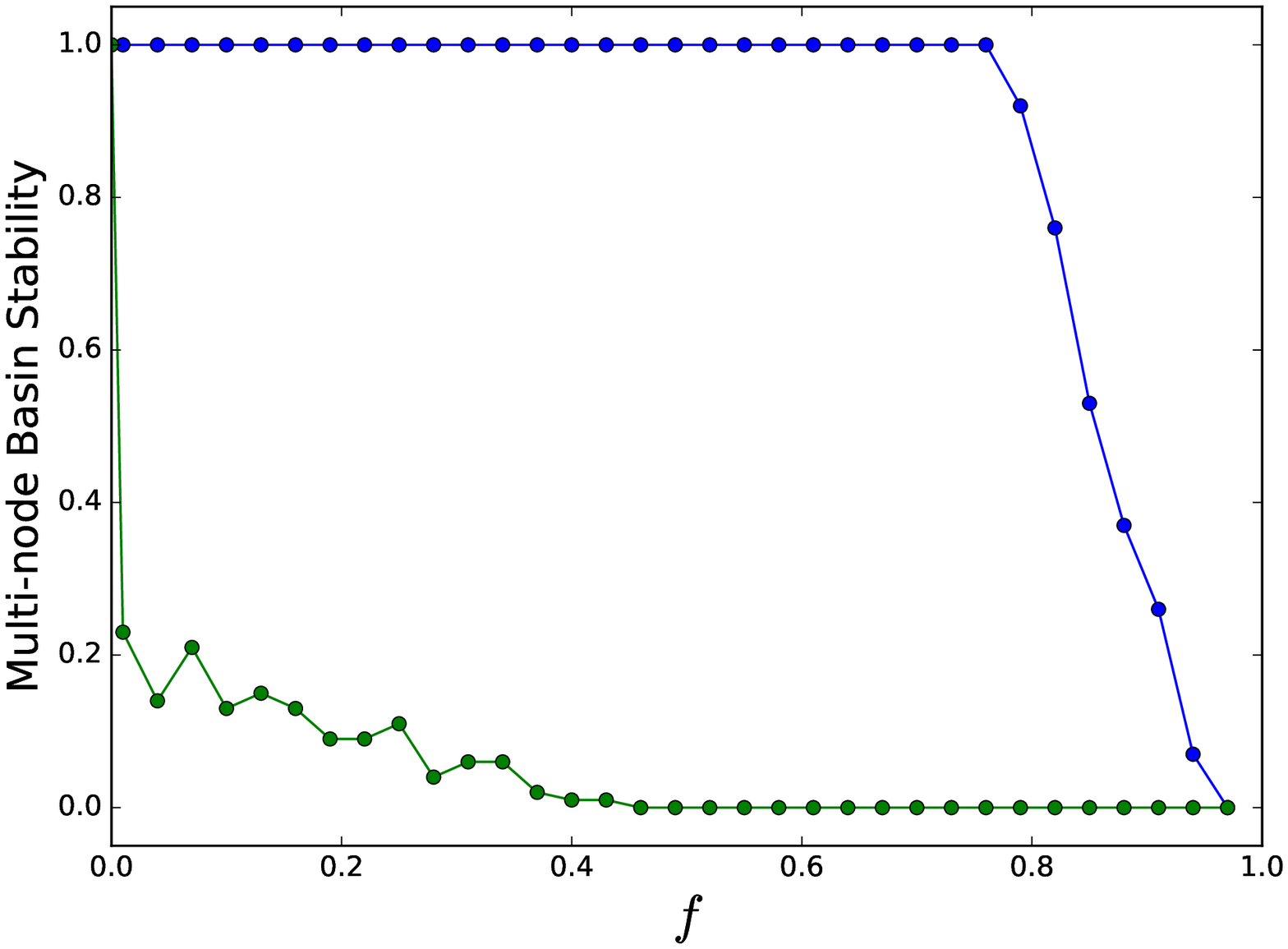}\\(b)

		\caption{Multi-node Basin Stability vs coupling strength for a Star network of size $N=100$: (a) the hub node is perturbed (blue) and a single peripheral node is perturbed (green); (b) Multi-node Basin Stability vs number of nodes perturbed in the Star network of bistable elements. Here the size of the network $N=100$ and coupling strength $C=1$. The blue curve represents the case where only peripheral nodes are perturbed, while green represents the case where the hub is perturbed alongwith peripheral nodes.}
		\label{bsvsc_star}
	\end{figure}

	Next we examine the multi-node basin stability of the network, for fraction $f$ of perturbed nodes ranging from $1/N$ (namely single node in Fig.~\ref{bsvsc_star}(a) ) to $f \sim 1$, namely the case where nearly all nodes in the system are perturbed. As evident from Fig.~\ref{bsvsc_star} (b), when only the peripheral nodes are perturbed, even for values of $f$ as high as $0.7$, there is no discernable difference in the basin stability, which remains close to $1$. This implies that even when more than half the nodes in the network are perturbed the entire system almost always recovers to the original state. In contrast, in Fig.~\ref{bsvsc_star}(a) shows the single-node basin stability for the case of the hub node being perturbed, where the basin stability is clearly drastically reduced and approaches zero very quickly. It is clear that just a {\em single} node is enough to destroy the stability of the network, if that node has very high degree, closeness and betweeness centrality, such as the hub node. These quantitative results are consistent with the qualitative spatiotemporal patterns observed in Fig.~\ref{Starspt}.

Further Fig. \ref{bsvsc_star}(b) shows the decline in Multinode Basin Stability with increasing fraction of perturbed nodes $f$. Interestingly, when only the peripheral nodes are perturbed, Basin Stability is close to one even when a very large fraction of nodes in the system are perturbed, and $f_{crit} \approx 0.7$. So when only peripheral nodes are perturbed, the perturbed nodes manage to return to their original state, {\em even when a majority of nodes in the system have been pushed to the basin of attraction of the other state}. However when the perturbed nodes include the central hub node, the basin stabilty declines rapidly with increasing $f$, and $f_{crit} \approx 1/N $. So our analysis reveals how significant the degree, closeness and betweeness of nodes are in determining the resilience of the network. In fact, very clearly, the {\em hub node holds the key to the maintenenance of the collective state}.

We can rationalize the above results from dynamics of the coupled system as follows: the influence of the neighbours on a particular node is through the local mean field generated by the nieghbouring nodes. Now it is clear if a node has few neighbours the influence of a perturbation on its neighbour will be very large. On the other hand, if a node is connected to many other nodes, as is typically true of nodes with high degree or betweenness centrality, such as a hub in a star network, the influence of peturbations on a node in its neighbourhood is scaled by a factor of $1/k$, where $k$ is the number of neighbours of the node. This implies that the effect of the peripheral nodes on the hub is much smaller than the hub on the periphery. Namely, the hub affects all peripheral nodes strongly with the coupling term being of $O(1)$. However a perturbation on a peripheral node affects the hub only through a coupling of $O(1/N)$, which is vanishingly small for large networks. Further the peripheral nodes do not affect each other directly, but only through perturbations propagating to the hub, while the hub  simultaneously affects all peripheral nodes.\\

{\bf Dynamics of a Random Scale-Free Network of Bistable Systems}\\

\begin{figure}[htb]
	\centering
	\includegraphics[width=\plotSize]{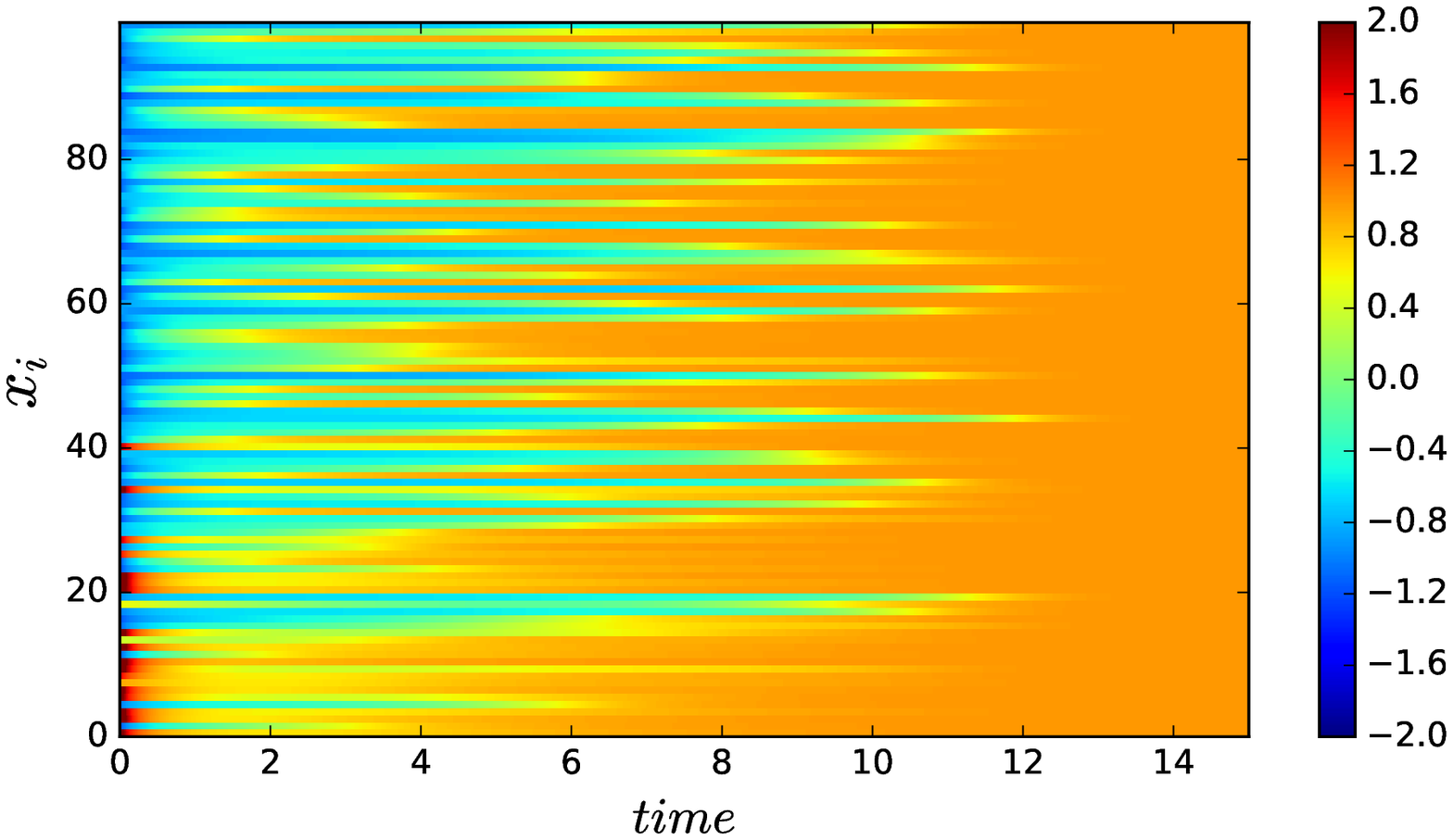}
	\includegraphics[width=\plotSize]{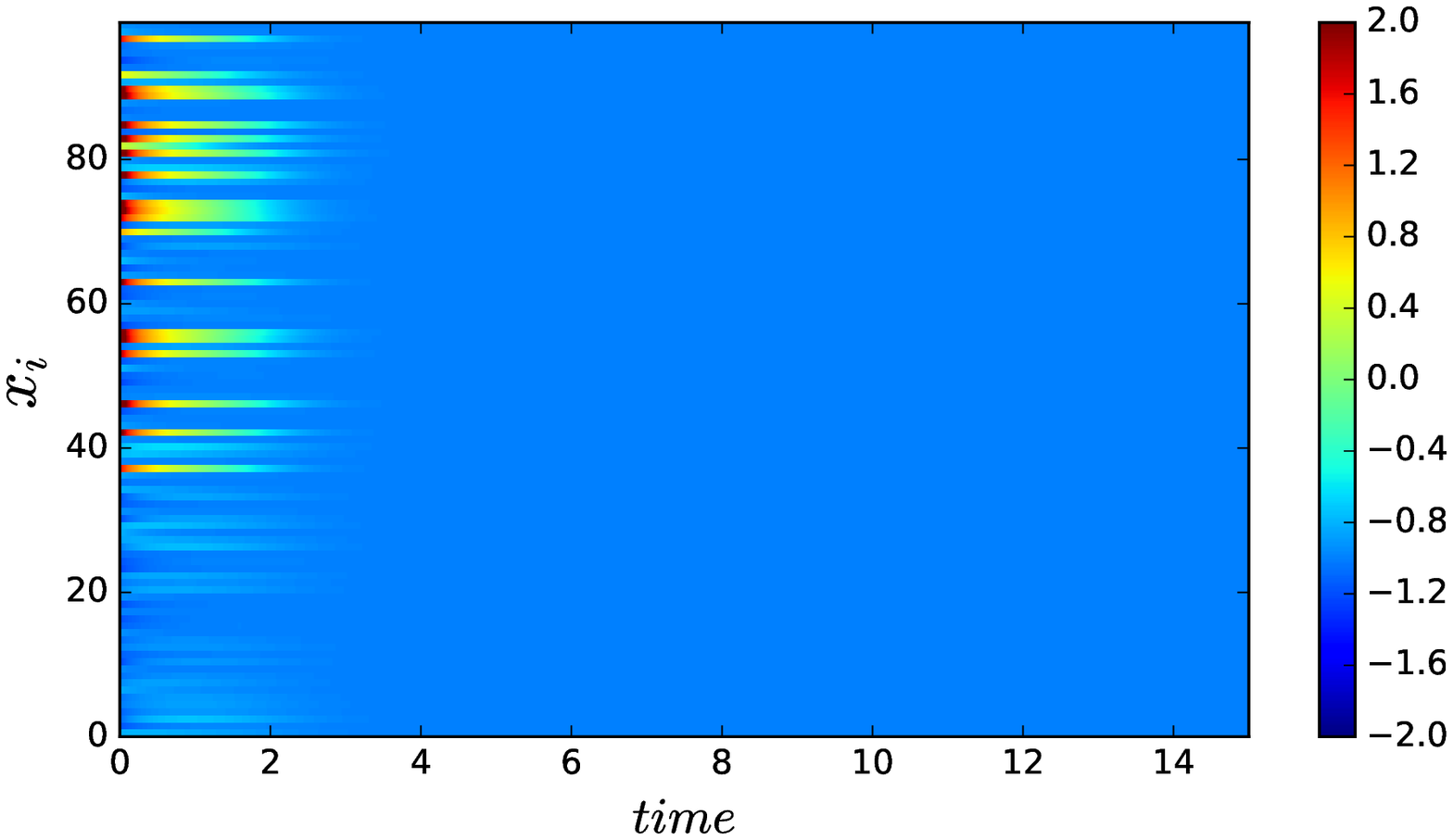}

\hspace{5cm} (a) \hfill (b) \hspace{5cm}
	\caption{Time evolution of $100$ bistable elements coupled in a Random Scale-Free network with $m=2$, given by Eqn.~\ref{main}, with coupling strength $C = 1$. In (a) $20$ nodes of highest betweeness centrality are perturbed; in (b)  $20$ nodes of lowest betweeness centrality are perturbed. Here the original state of the networks has all nodes in the negative well, i.e. all $x_i < 0$.}
	\label{RSFspt}
\end{figure}

We will now go on to explore Random Scale-Free (RSF) Networks of bi-stable dynamical elements. In particular, we construct these networks via the Barabasi-Albert preferential attachment algorithm, with the number of links of each new node  denoted by parameter $m$ \cite{scalefree}. The network is characterised by a fat-tailed degree distribution. Specifically we will display results for networks of size $N=100$, with $m=1$ and $m=2$. Figs.~\ref{RSFspt}a-b shows two contrasting representative cases where (a) twenty nodes with highest betweeness centrality are perturbed, and (b) twenty nodes with the lowest betweeness centrality were perturbed. It is observed that perturbation on nodes with high betweeness centrality destabilises the entire network, and the perturbed nodes rapidly drag all the other nodes to a different well. This is evident from the switched colors of the asymptotic state in Fig.~\ref{RSFspt}a. On the other hand, when the perturbed nodes have low betweeness centrality, the network recovers quickly from the perturbation and reverts to the original well, as clearly seen in Fig.\ref{RSFspt}b. These completely different outcomes occur even though the {\em number of perturbed nodes is the same} in both cases, thereby clearly illustrating that nodes with high betweeness-centrality have much stronger influence on the global stability of the system.

We will now present the dependence of the global stability of the collective dynamics on different centrality measures in this heterogeneous network, quantitatively, through multi-node basin stability measures. In particular, in order to explore the correlation between a given centrality measure of the nodes and the resilience of the system, we will estimate the multi-node basin stability under perturbations on sub-sets of nodes with increasing (or decreasing) values of the centrality under consideration. That is, we order the nodes according to the centrality we are probing, and consider the effect of perturbations on fraction $f$ of nodes with the highest (or lowest) centrality. 

\begin{figure}[htb]
	\centering
	\includegraphics[width=0.3\linewidth]{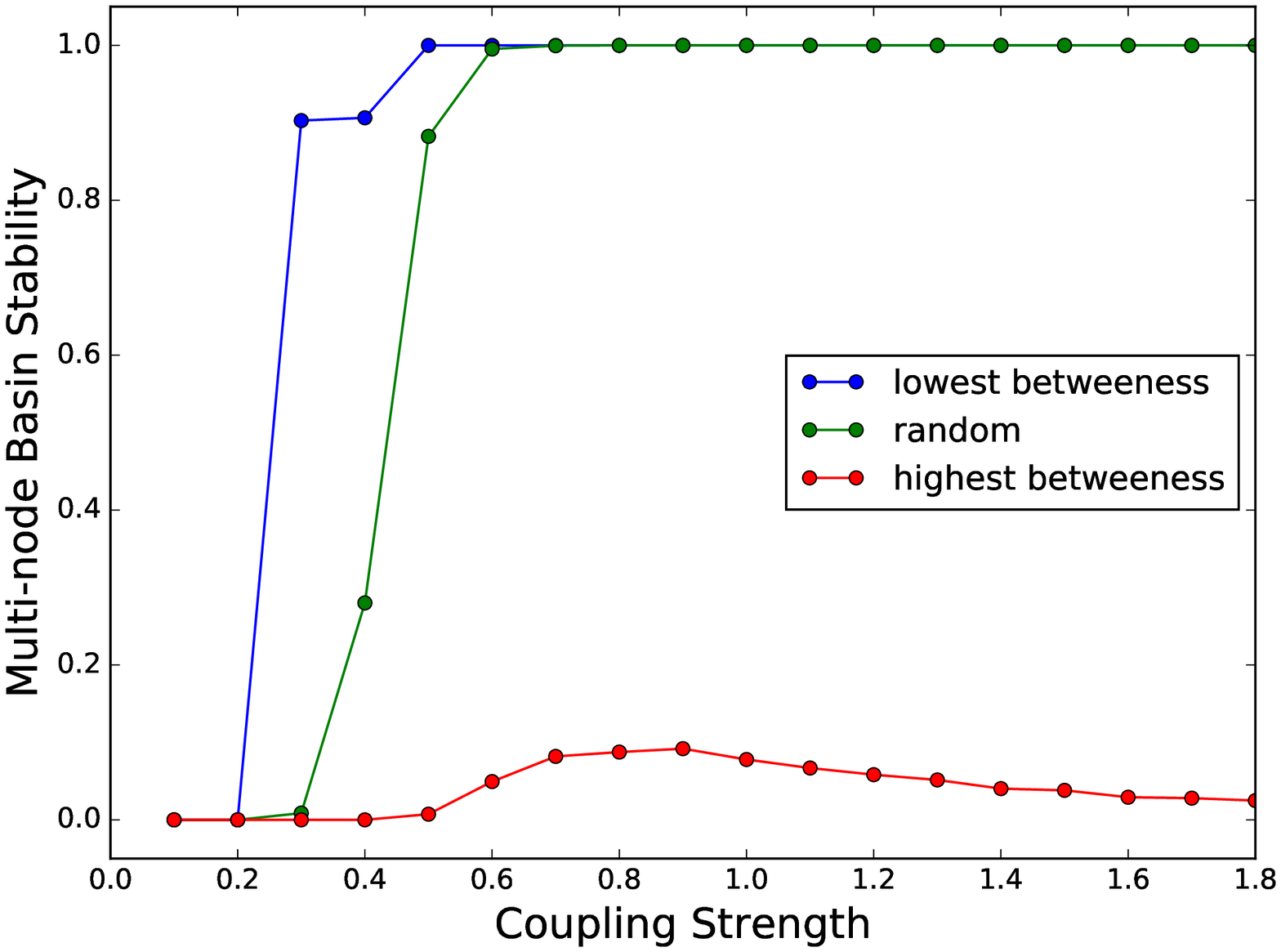}
	\includegraphics[width=0.3\linewidth]{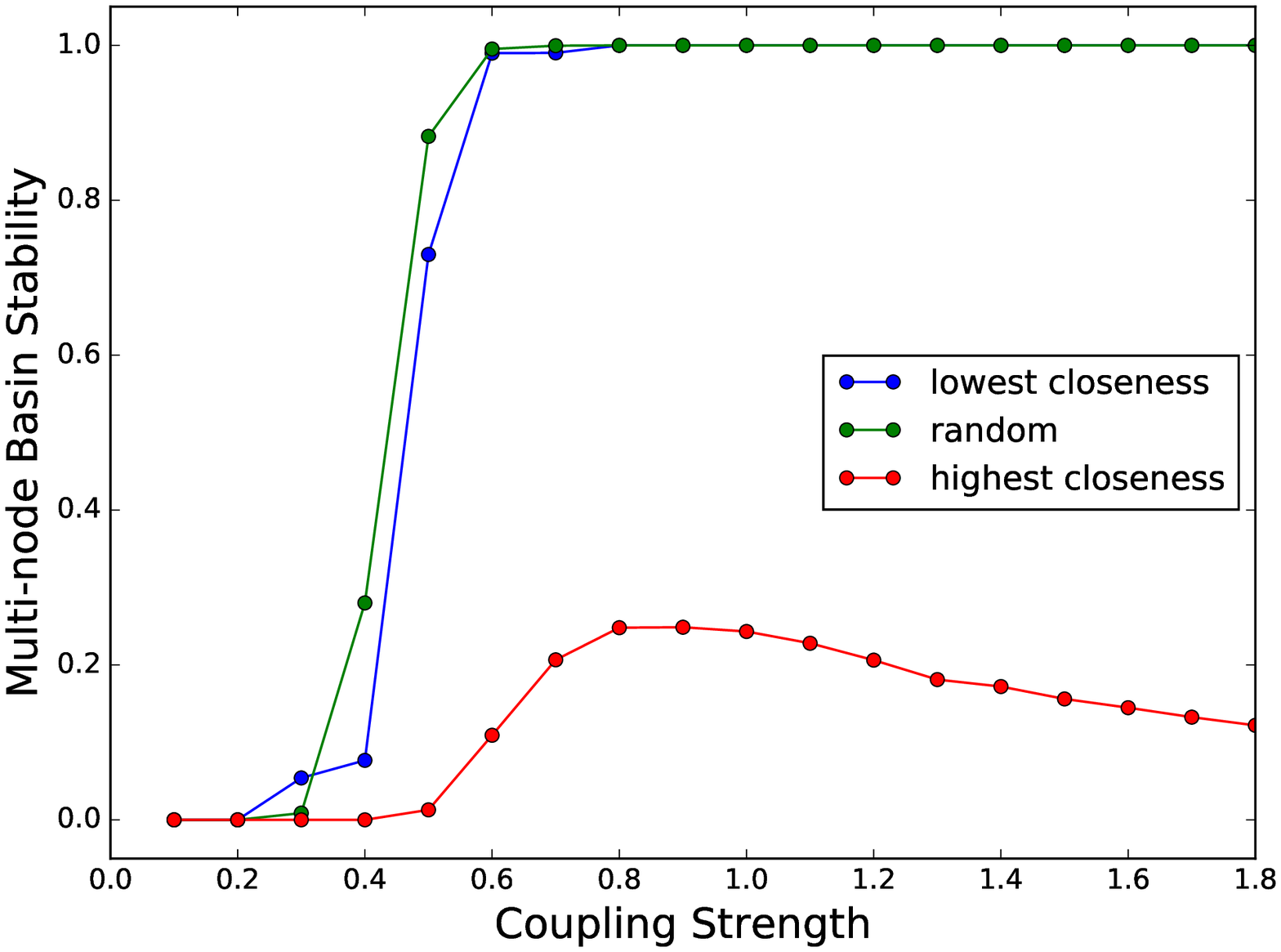}
	\includegraphics[width=0.3\linewidth]{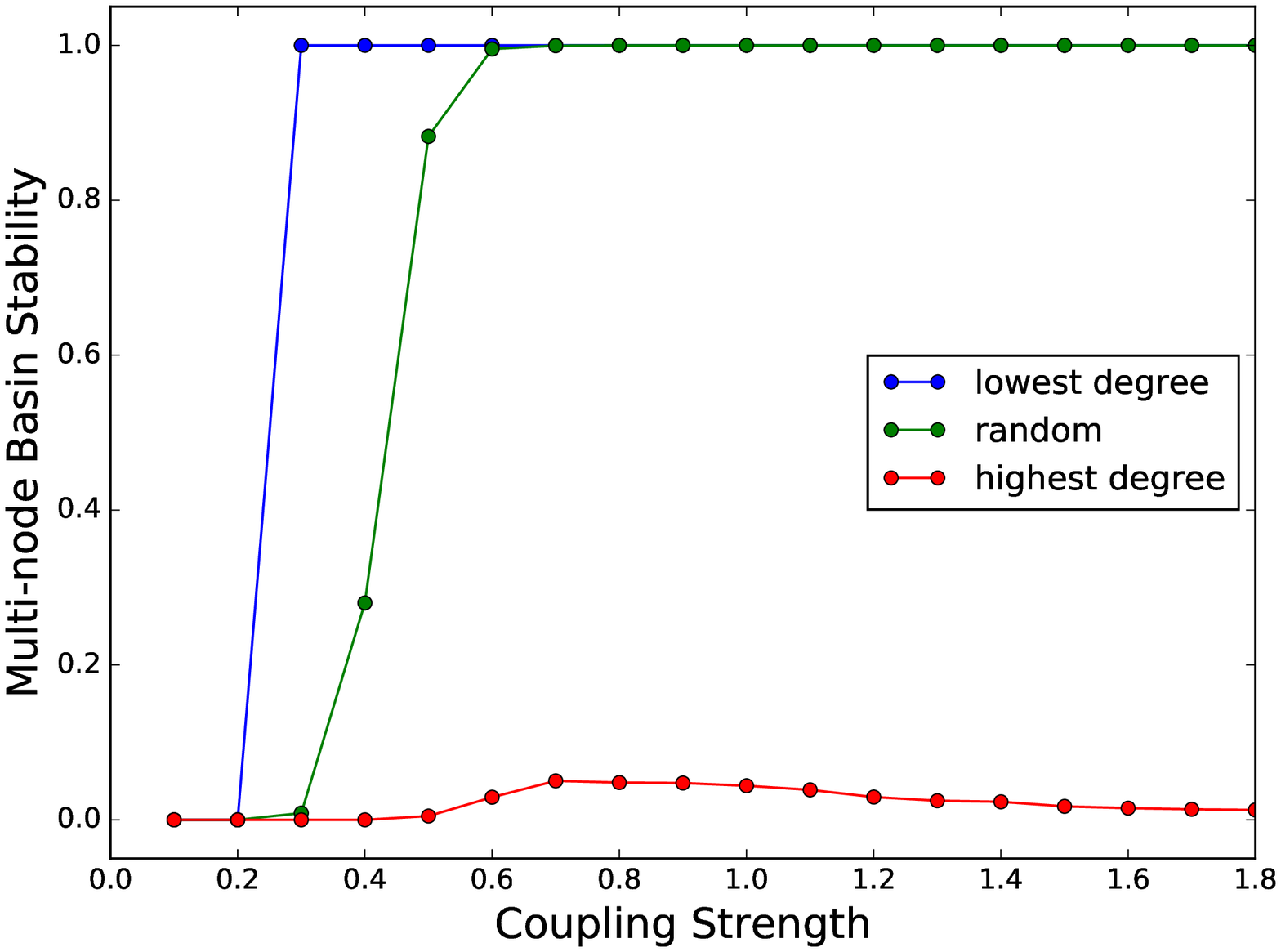}

	\includegraphics[width=0.3\linewidth]{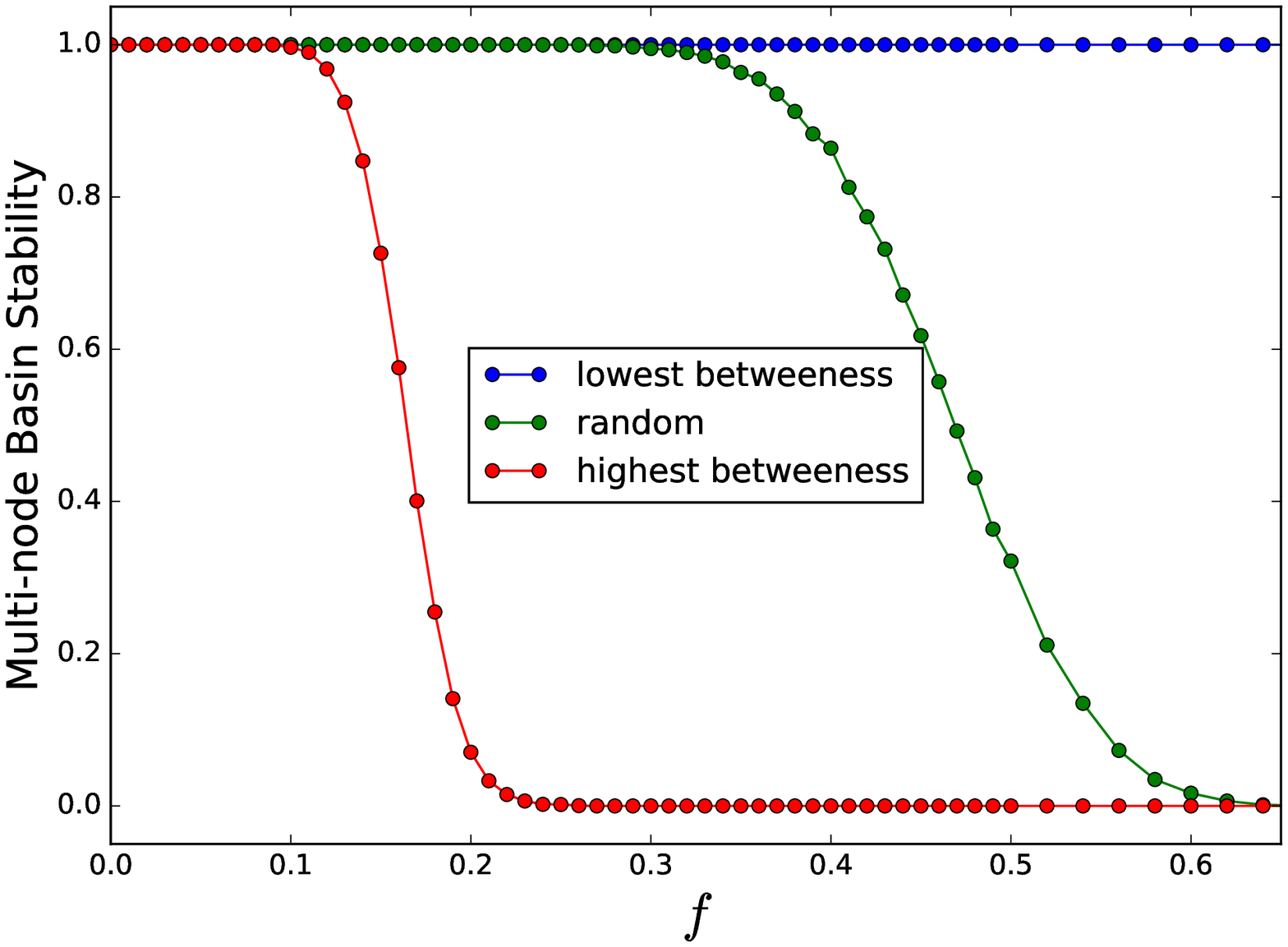}
	\includegraphics[width=0.3\linewidth]{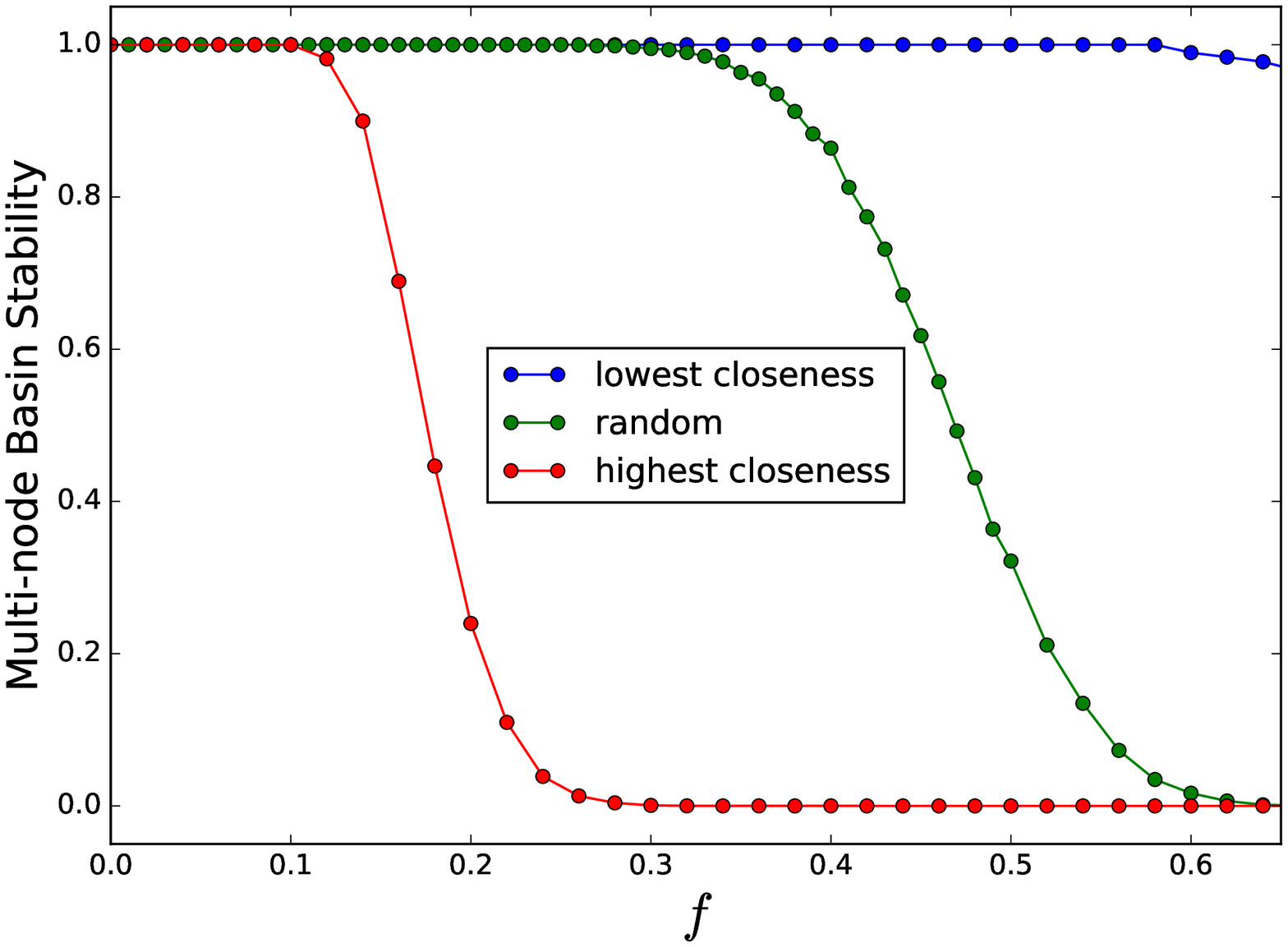}
	\includegraphics[width=0.3\linewidth]{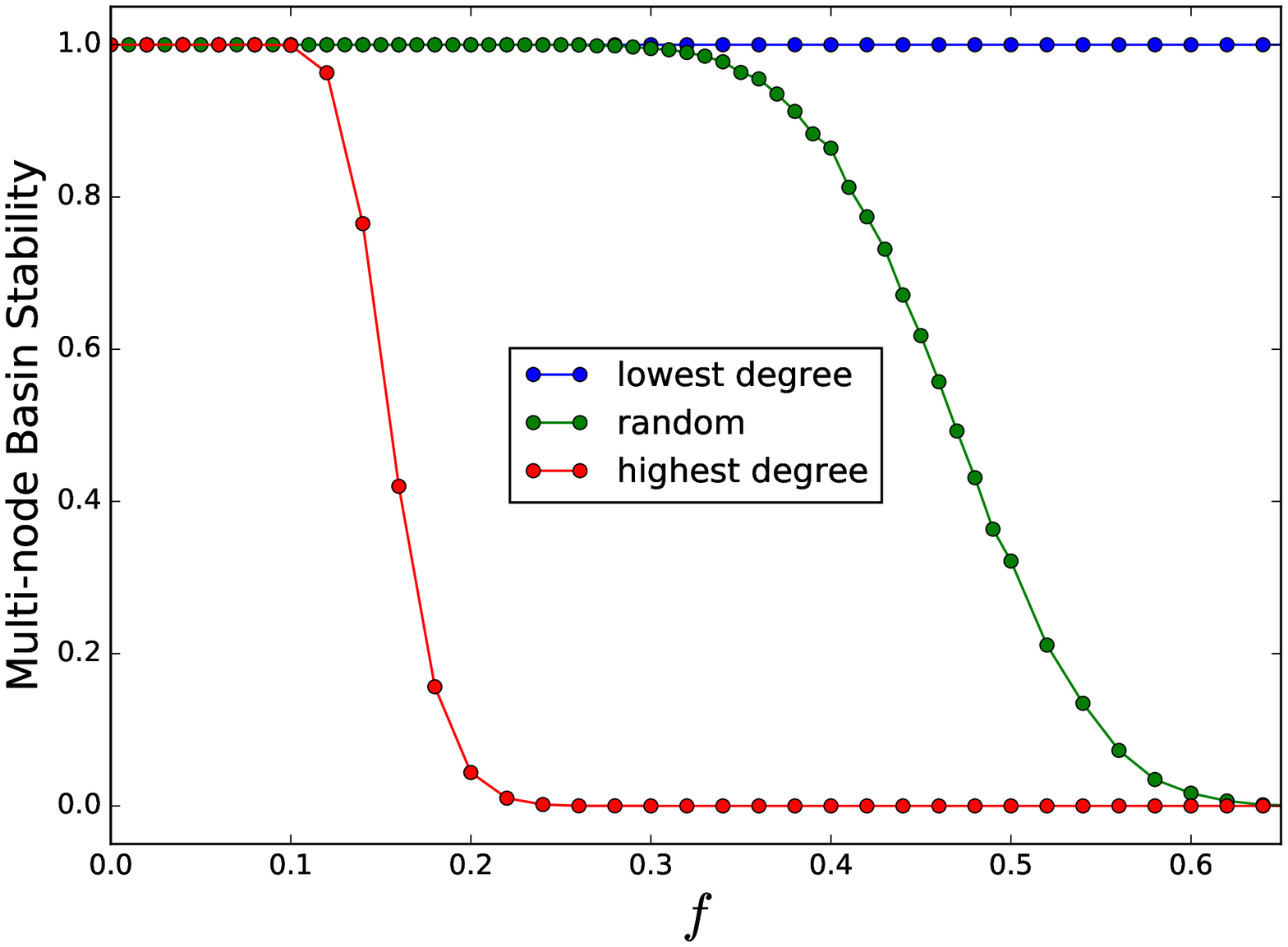}

\hspace{2cm} (a) \hfill (b) \hfill (c) \hspace{2cm}
	\caption{(a) Dependence of the Multinode Basin Stability of Random Scale-Free networks of size $N=100$, with $m=2$, on coupling strength $C$, with $f=0.2$ (top panels) and fraction $f$ of perturbed nodes, with $C=1$ (bottom panels). In the panels, three cases are shown. In the first case, the perturbed nodes are chosen at random (green curves). In the second case (red curves) the perturbed nodes are chosen in descending order of (a) betweeness centrality, (b) closeness centrality and (c) degree (i.e. the perturbed nodes are the ones with the highest $b$, $c$ or $k$ centrality measures). In the third case (blue curves) the perturbed nodes are chosen in ascending order of (a) betweeness centrality, (b) closeness centrality and (c) degree (i.e. the perturbed nodes are the ones with the lowest $b$, $c$ or $k$ centrality measures).}
	\label{rsf1}
\end{figure}

The influence of perturbations on nodes with the highest and lowest betweeness, closeness and degree centrality in a Random Scale-Free network are displayed in Figs.~\ref{rsf1}a-c. The broad trends are similar for all three centrality measures, and it is clearly evident that when nodes with the highest betweeness, closeness and degree centrality are perturbed, multi-node basin stability falls drastically. On the other hand, perturbing the same number of nodes of low centrality leaves the basin stability virtually unchanged. Further, when nodes of low centrality are perturbed, for sufficiently high coupling strengths, the network almost always recovers to its original state, yielding a basin stability of $1$. So one can conclude that perturbing nodes with high betweeness, closeness and degree centrality destroys the synchronized state readily, while perturbing nodes of low centrality allows the perturbed nodes to return to the original state, thereby restoring the sychronized state. For reference, Fig.~\ref{rsf1} also shows the basin stability of a network where the perturbed nodes are randomly chosen, corresponding to {\em random attacks} on a subset of nodes. Clearly, a {\em targetted attack} on nodes with high centrality can destroy the collective dynamics much more efficiently than random attacks.

\begin{figure}[htb]
	\centering
	\includegraphics[width=0.3\linewidth]{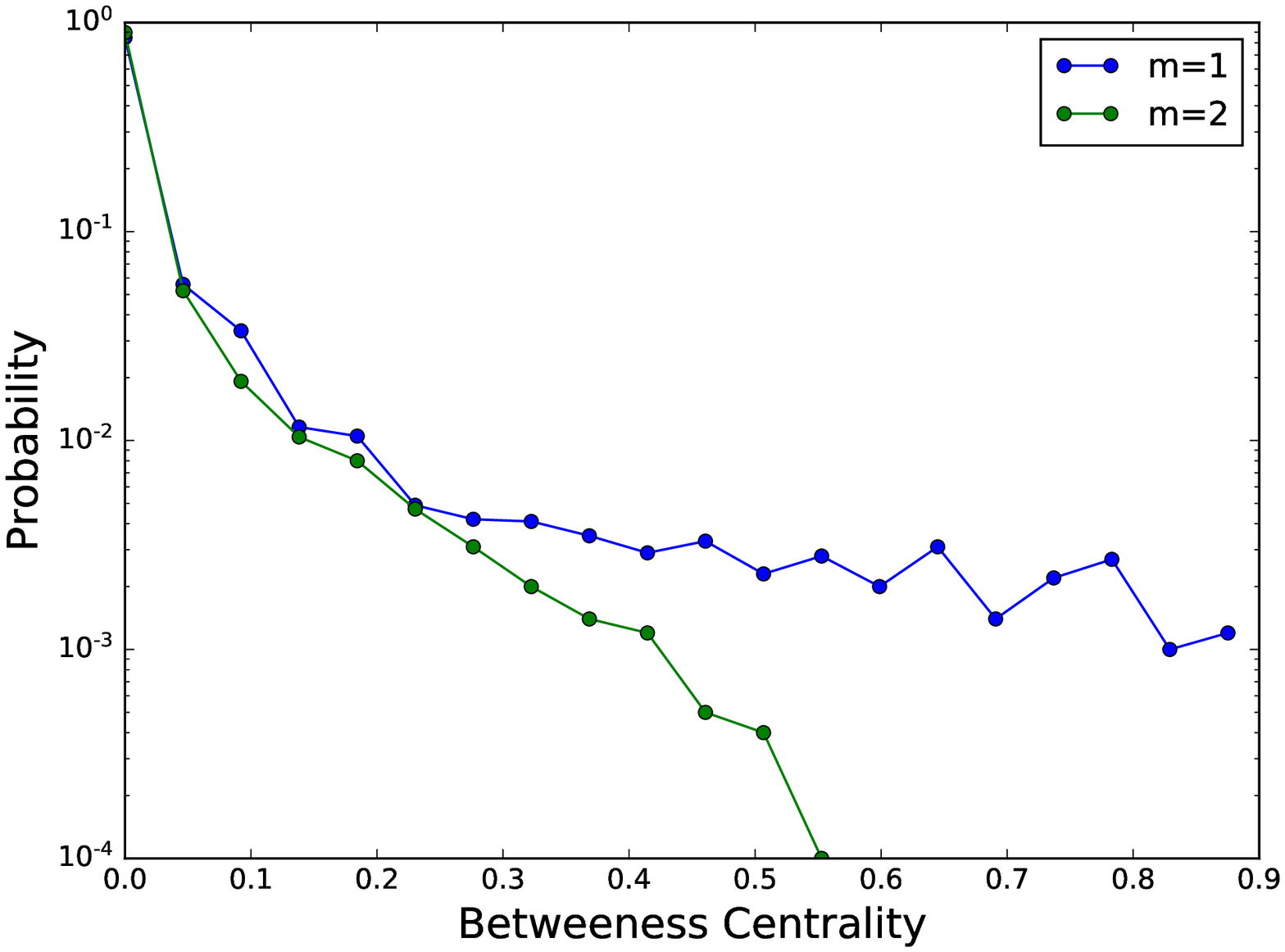}
	\includegraphics[width=0.3\linewidth]{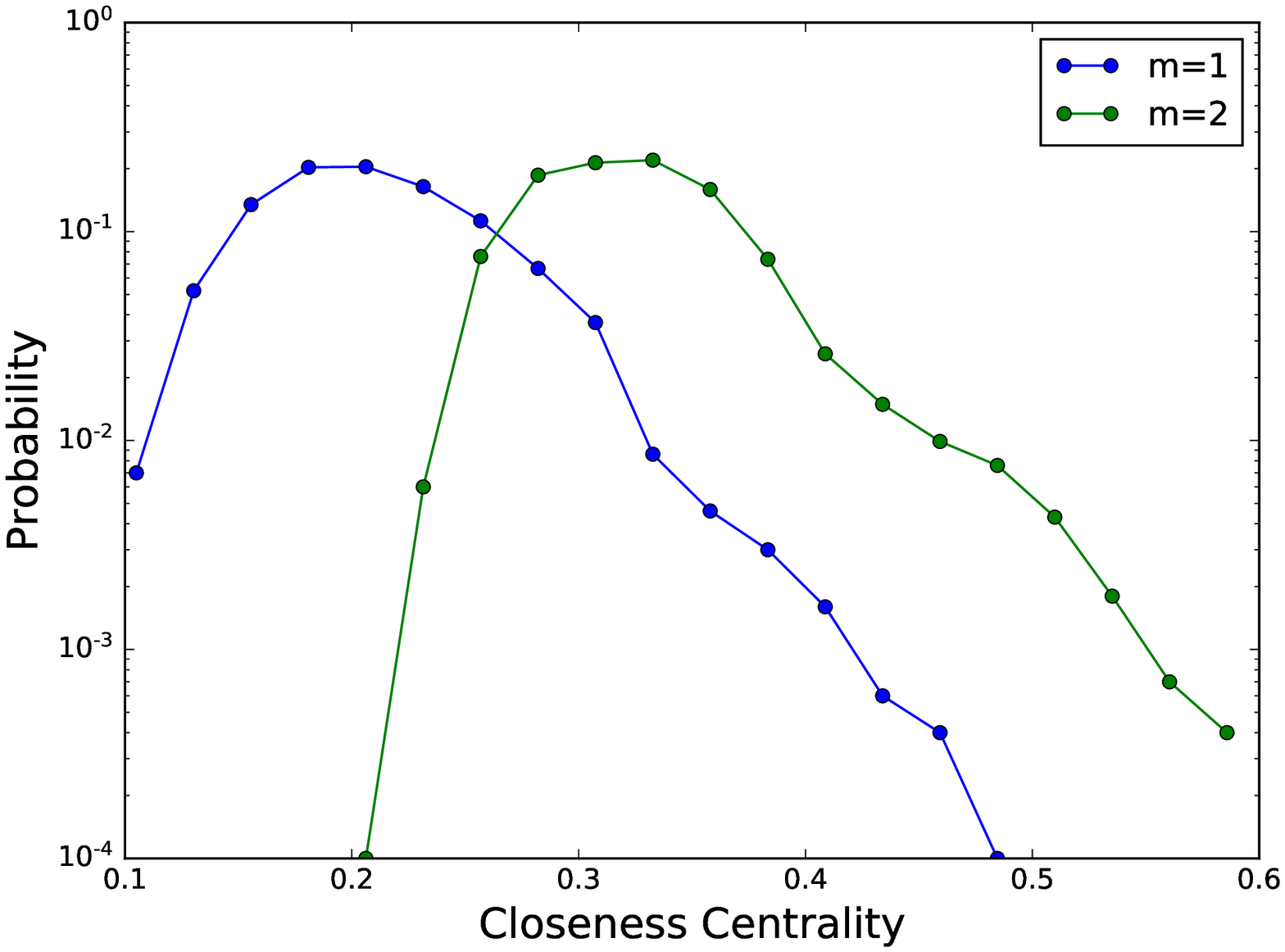}
	\includegraphics[width=0.3\linewidth]{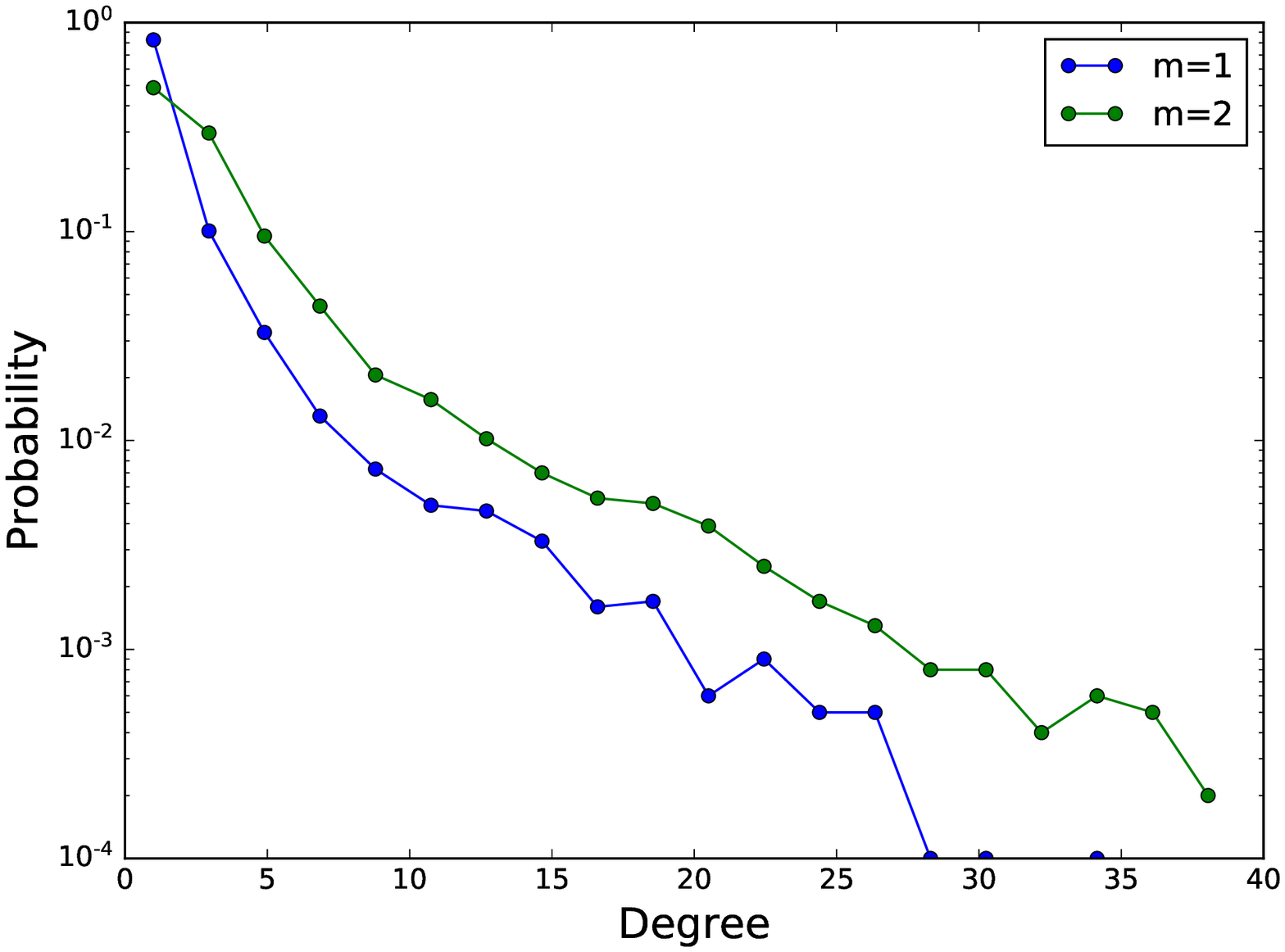}

	\includegraphics[width=0.3\linewidth]{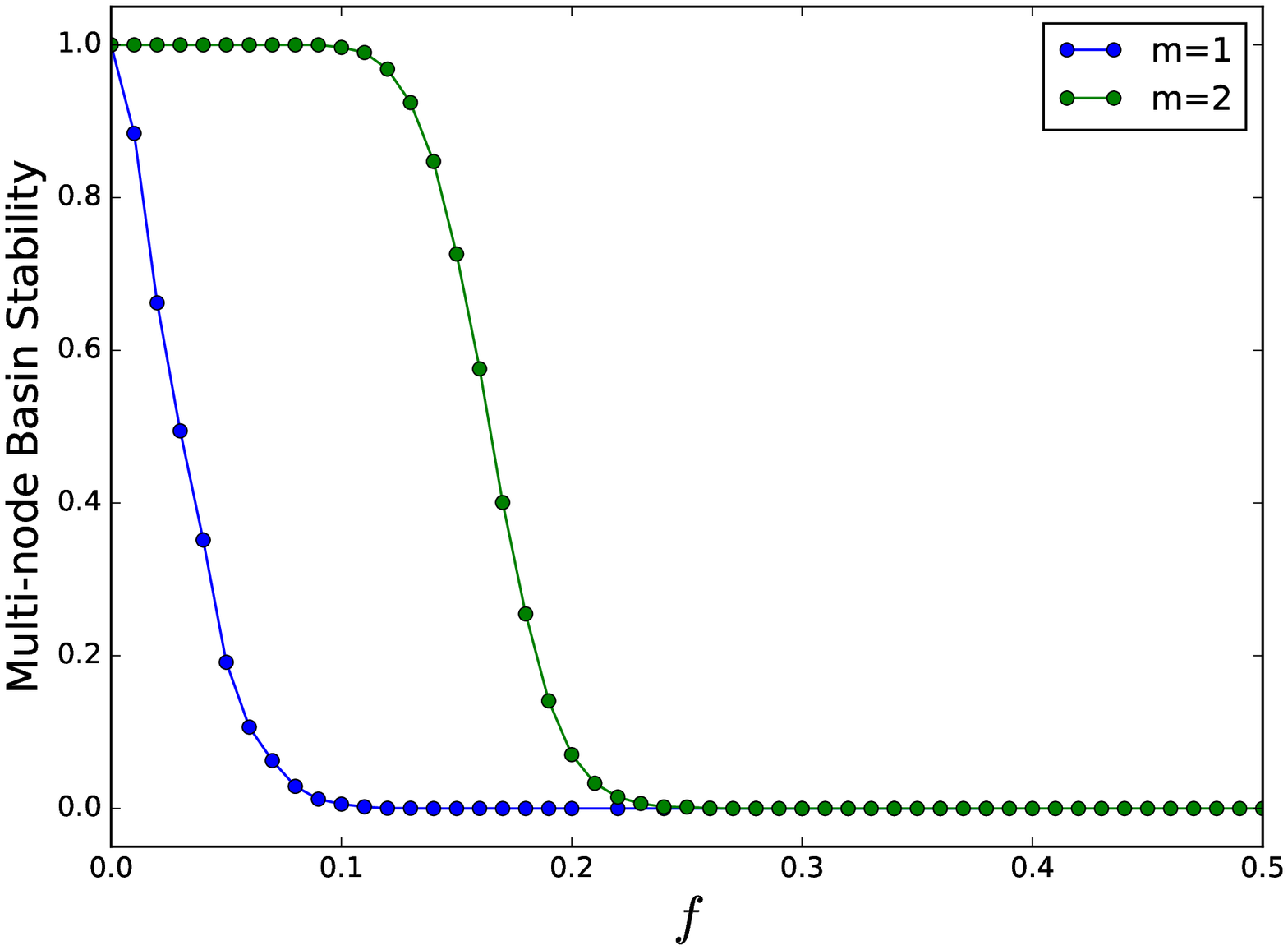}
	\includegraphics[width=0.3\linewidth]{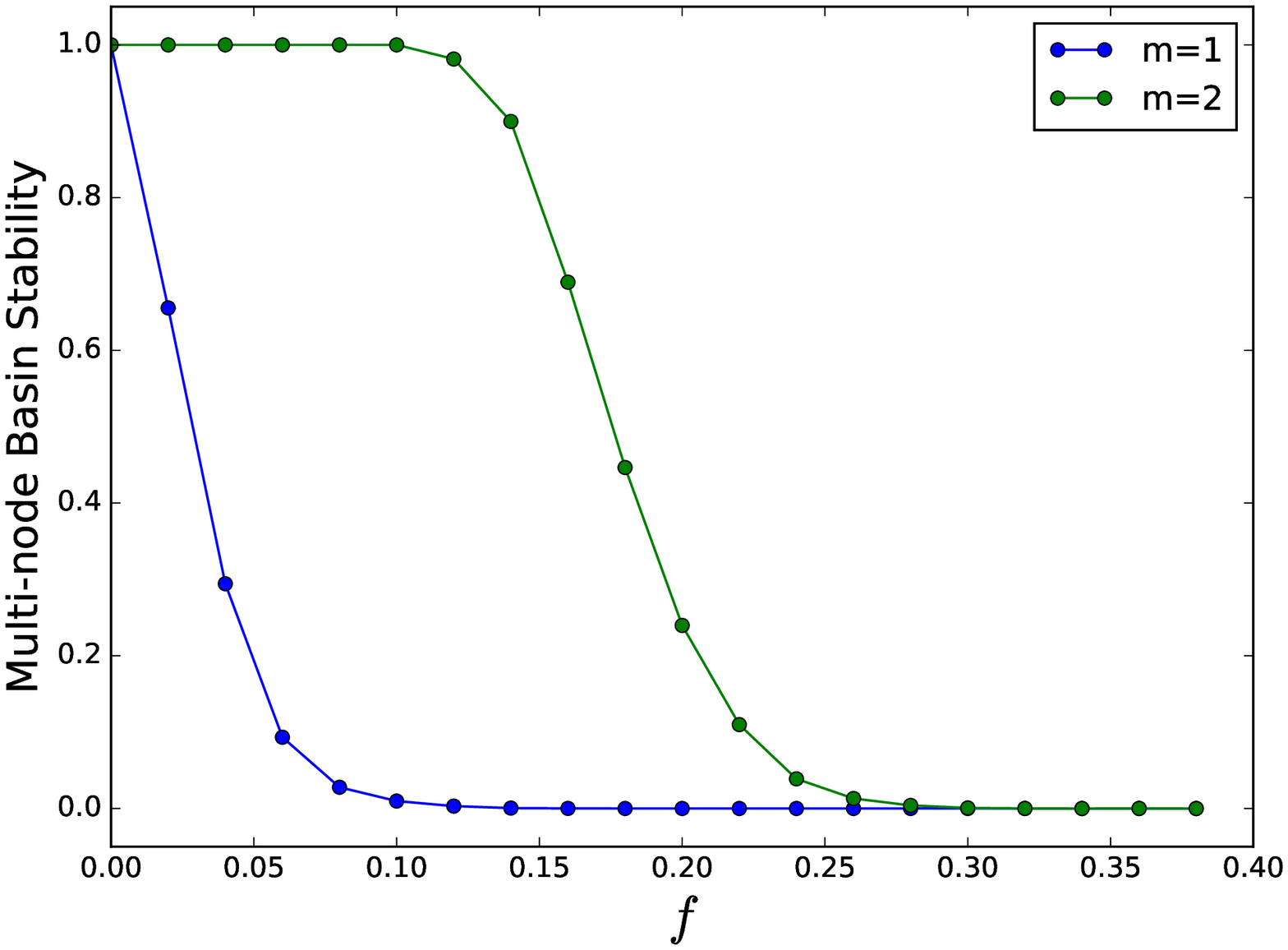}
	\includegraphics[width=0.3\linewidth]{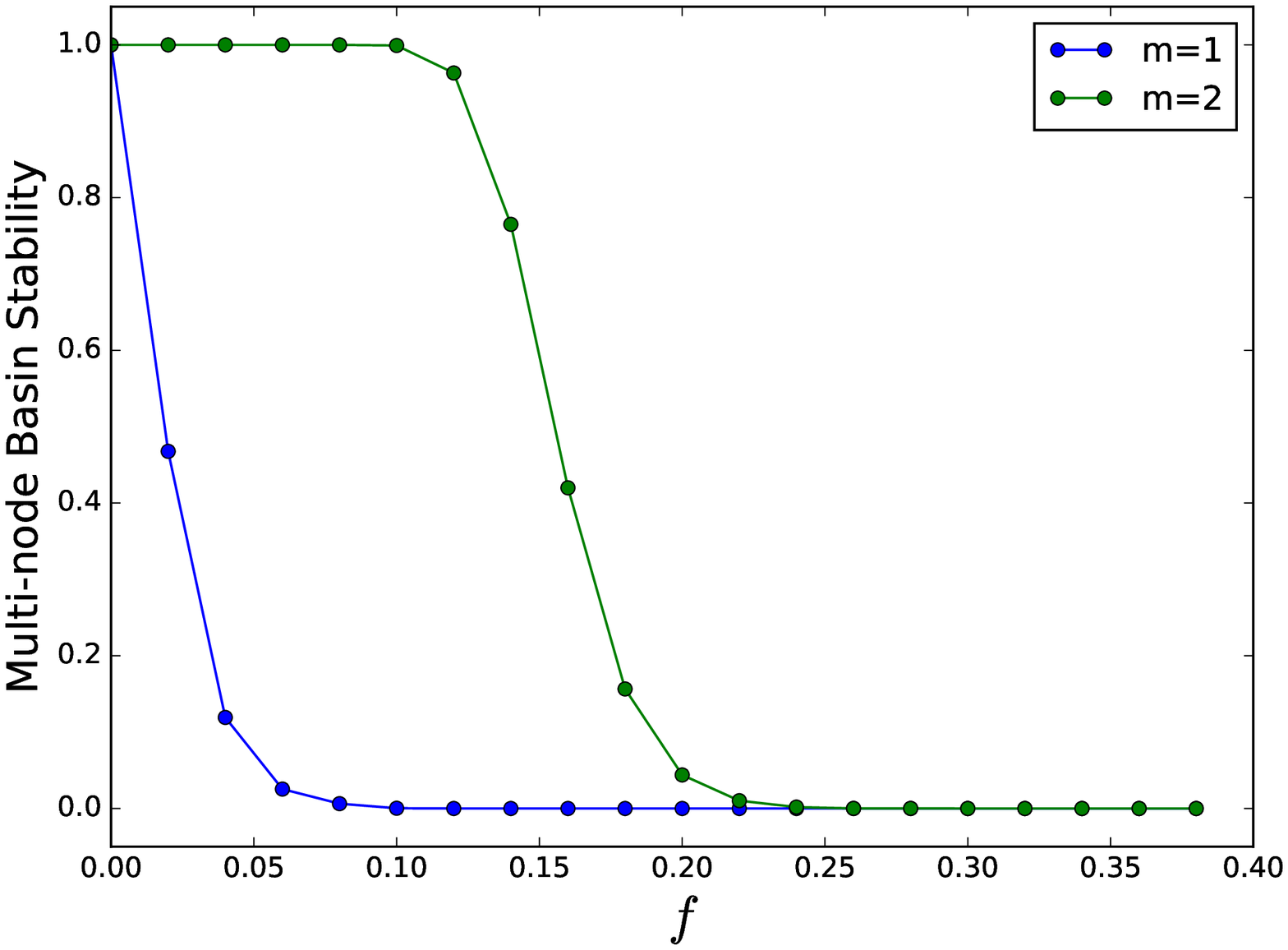}

\hspace{2cm} (a) \hfill (b) \hfill (c) \hspace{2cm}

	\caption{Top panel: Probability distribution of the (a) betweeness centrality, (b) closeness centrality and (c) degree of the nodes in a Random Scale-Free network of size $N=100$, with $m=1$ (blue) and $m=2$ (green). Bottom panel:  Multi-node basin Stability vs fraction $f$ of nodes perturbed, for Random Scale-Free network of size $N=100$, coupling strength $C=1$, with $m=1$ and $m=2$, where the perturbed nodes are chosen in descending order of (a) betweeness centrality, (b) closeness centrality and (c) degree (i.e. the perturbed nodes are the ones with the highest $b$, $c$ or $k$ centrality measures).}
	\label{m12_prob}
\end{figure}

Now we investigate which centrality measure is most crucial in determining the global robustness of collective behaviour in the network. We do this through the following numerical experiment: we compare the basin stability of the collective dynamics of Random Scale-Free networks with $m=1$ and $m=2$. Interestingly, the distribution of the betweeness centrality, closeness centrality and degree of the nodes in RSF networks with $m=1$ and $m=2$ are significantly different, as evident in Fig.~\ref{m12_prob} (top panels). It is clear that for $m=1$ the distribution of the degree and closeness centrality of the nodes in the network is shifted towards lower $k$ and $c$ values as compared to RSF networks with $m=2$, while the distribution of betweeness centrality shifts towards higher values in RSF networks with $m=1$ vis-a-vis the distribution of the betweeness centrality in RSF networks with $m=2$. So in RSF networks  with $m=1$ the nodes with the highest betweeness centrality typically have significantly higher $b$ than in RSF networks with $m=2$ of the same size. On the other hand, since the tail of the probability distribution of degree and closeness centrality of the nodes in a RSF network with $m=2$ extends further than that in a RSF network with $m=1$, the nodes with the highest degree and closeness centrality typically have lower $k$ and $c$ in RSF networks with $m=1$ compared to RSF networks with $m=2$. So these networks can potentially {\em provide a test-bed for determining which of the centrality properties most crucially influence dynamical robustness}. Note that it was not possible to use the Ring to probe this issue, as all nodes there have identical centrality propoerties. Nor did the Star network offer a system where one could distinguish between the effects of different centrality measures on dynamical robustness, as the nodes there split into two classes, the single hub and the periphery, with all the peripheral nodes having identical betweeness, closeness and degree. However, one can compare the response of Random Scale-Free networks with different $m$ to probe which nodal property renders a heterogeneous network most vulnerable to large localized perturbations.


Fig.~\ref{m12_prob} (bottom panels) displays the dependence of the multi-node Basin stability on the fraction of perturbed nodes $f$ in the Random Scale-Free network with $m=1$ and $m=2$. As number of nodes perturbed increases, the multi-node basin stability falls significantly for RSF networks with $m=1$, while RSF networks with $m=2$ remains robust up to a critical fraction $f_{crit}$ of perturbed nodes, with $f_{crit} \sim 0.2$. One can rationalize this, by noting the difference in the typical values of betweeness centrality at the highest end in the RSF network with $m=1$ and $m=2$. For instance, if one considers $10 \%$ of nodes with the highest betweeness centrality in these networks of size $N=100$, typically $b$ lies between $0.1$ to $0.9$ for $m=1$ and between $0.01$ and $0.5$ for $m=2$. So the marked difference in the sensitivity of the global stability to perturbations in Random Scale-Free networks with $m=1$ and $m=2$ stems from the higher betweeness centrality of the nodes in the former network.

Now when nodes of the highest closeness centrality and degree are perturbed we observe the same trend as above. This occurs inspite of the tail of the distribution of closeness centrality and degree extending to higher values for RSF networks with $m=2$ as compared to RSF networks with $m=1$, implying that the nodes with highest degree and closeness centrality for the $m=2$ case will have a larger value of $k$ and $c$, as compared to the $m=1$ case. So one may have expected that the RSF network with $m=2$ would be less stable than the RSF network with $m=1$. However, the observations are contrary to this expectation and this surprising result stems from the following: the set of nodes with the highest betweeness centrality, closeness centrality and degree, overlap to a very large extent. So for instance, for $f=0.1$ in a network of size $N=100$, the set of $10$ nodes with the highest betweeness centralities, is practically the same as the set of nodes with the highest closenes centralities and highest degrees. However, in the RSF network with $m=1$ these nodes have higher betweeness centrality, while having lower closeness centrality and degree, than the corresponding set in the RSF network with $m=2$. Now higher betweeness centrality should inhibit stability, while lower closeness and degree should aid the stability of the collective state. So the comparative influence of these two opposing trends will determine the comparative global stability of these two classes of networks. If the betweeness centrality of the perturbed nodes is more crucial for stability, the multi-node Basin Stability of the network with $m=1$ will go to zero faster than the network with $m=2$. On the other hand if closeness centrality (and/or degree) of the perturbed nodes dictates global stability rather than betweeness centrality, the network with $m=2$ will lose global stability faster than the one with $m=1$. Now, since we find that network with $m=1$ always loses stability faster than the network with $m=2$, we can conclude that the {\em effect of betweeness centrality on the global stability is more dominant than the effect of the closeness centrailty and degree of the perturbed nodes}.

\begin{figure}[htb]
	\centering
	\includegraphics[width=\plotSize]{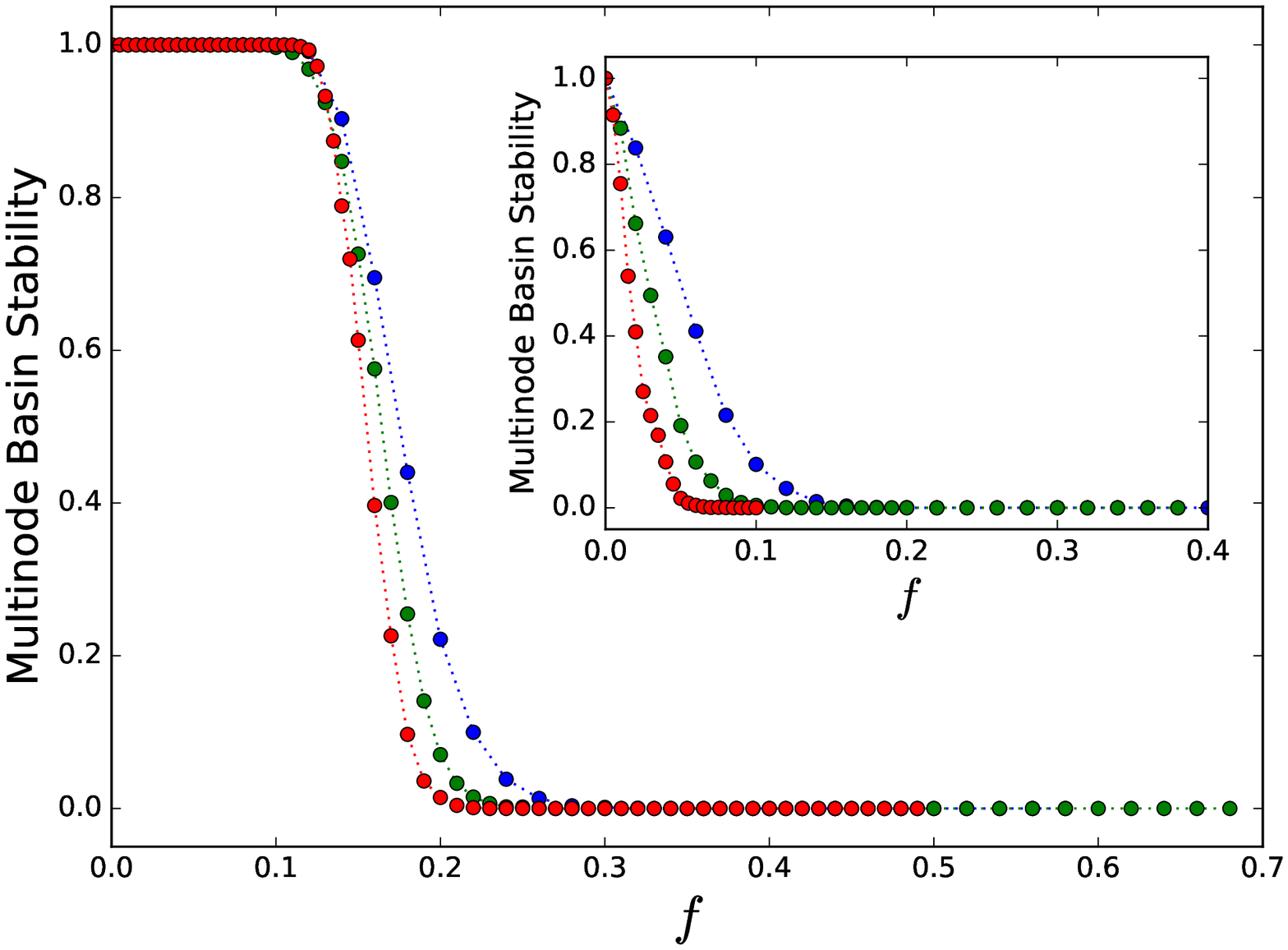}
	\includegraphics[width=\plotSize]{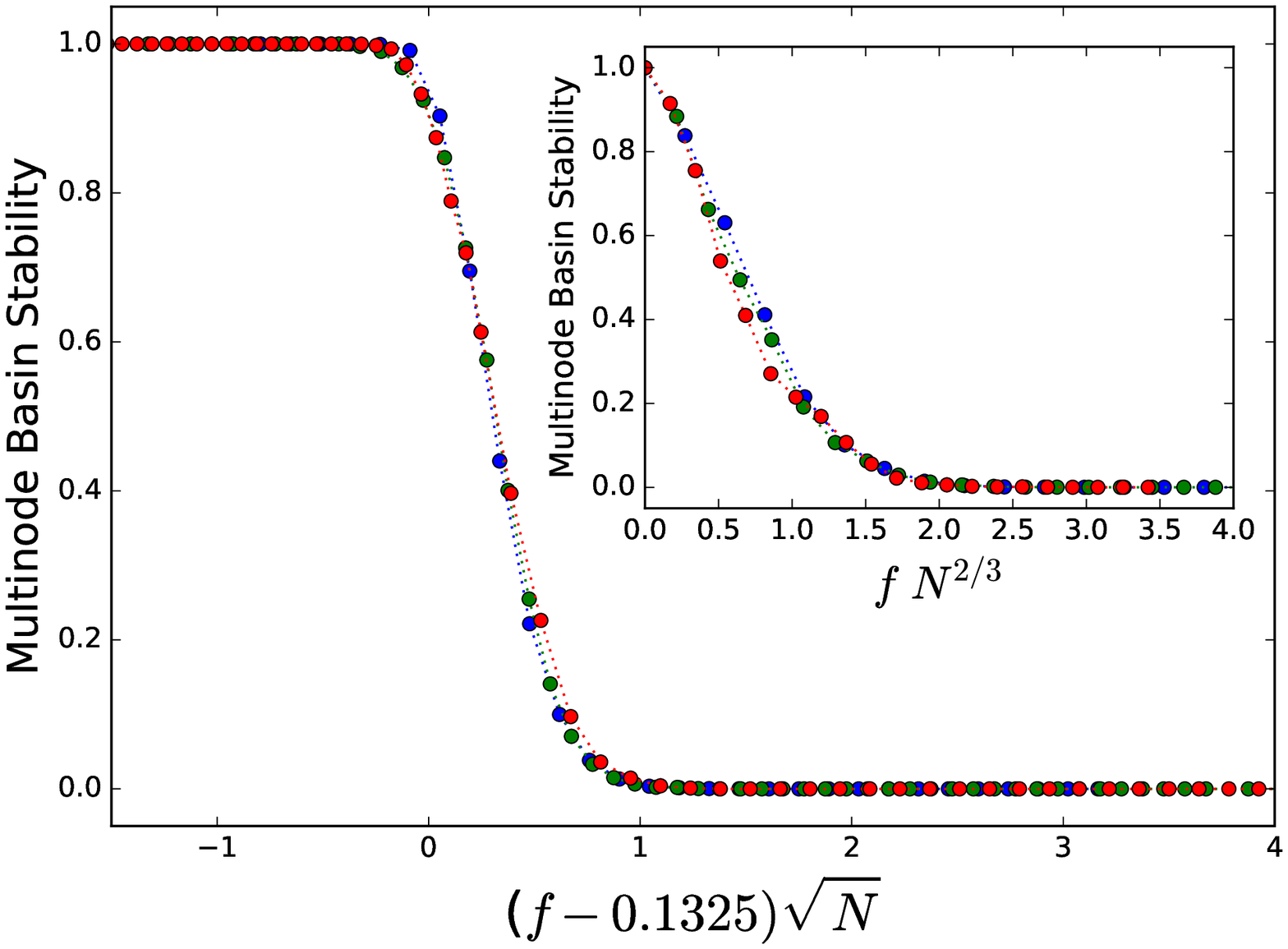}

\hspace{5cm} (a) \hfill (b) \hspace{5cm}
	\caption{(a) Multi-node basin Stability vs fraction $f$ of nodes perturbed, for Random Scale-Free network of size $N=50$ (blue), $100$ (green), $200$ (red) for $m=2$ and $m=1$ (inset). (b) Scaling resulting in data collapse, for the case of $m=2$ and $m=1$ (inset). The nodes perturbed are the ones with highest value of betweeness centrality.}
	\label{bsvsf_rsf_varyN}
\end{figure}

Lastly, we study the effect of system size on multi-node basin stability, perturbing nodes in decreasing order of betweeness centrality. Figs.~\ref{bsvsf_rsf_varyN}a-b shows the results for networks sizes ranging from $50$ to $200$. We have found an appropriate finite-size scaling that allows data collapse (cf. Fig.~\ref{bsvsf_rsf_varyN} insets), and this indicates the value $f_{crit}$ in the limit of large network size. We observe that a Random Scale-Free network with $m=1$ yields $f_{crit} \rightarrow 0$ (i.e. the smallest fraction of perturbed nodes destroy the collective state), while $f_{crit} \sim 0.2$ for the case of $m=2$. So a RSF network with $m=2$ is more robust to localized perturbations than a a RSF network with $m=1$, as in the $m=2$ case, even when nearly $20 \%$ of the nodes of the highest betweeness centrality are perturbed the entire network still manages to return to the original state. This compelling difference again arises due to the fact that the highest betweeness centrality found in the RSF network with $m=1$ is significantly higher on an average than that in RSF networks of the same size with $m=2$. This again corroborates the results in Fig.~\ref{m12_prob}, and highlights the profound influence of betweeness centrality on global stability.\\

{\bf Robustness of the phenomena:} \\

In order to ascertain the generality of our observations, we have considered different nonlinear functions $F(x)$ in Eq.(1). For example, we explored a system of considerable biological interest, namely, a system of coupled synthetic gene networks.  We used the quantitative model, developed in \cite{hasty}, describing the regulation of the operator region of $\lambda$ phase, whose promoter region consists of three operator sites. The chemical reactions describing this network, given by suitable re-scaling yields \cite{hasty} 
$$F_{gene}(x) = \frac{m (1 + x^2 + \alpha \sigma_1  x^4)}
				{1 + x^2 + \sigma_1 x^4 + \sigma_1 \sigma_2 x^6} - \gamma_x x$$
where $x$ is the concentration of the repressor. The non linearity in this $F_{gene} (x)$ leads to a double well potential, and different $\gamma$ introduces varying degrees of asymmetry in the potential. We studied a system of coupled genetic oscillators given by: $\dot{x_i} = F_{gene}(x_i) + C (\langle x_i^{nbhd} \rangle -x_i)$, where $C$ is the coupling strength and $\langle x_i^{nbhd} \rangle$ is the local mean field generated by the set of neighbours of site $i$. 

Further we studied different networks of a piece-wise linear bi-stable system, that can be realised efficiently in electronic circuits \cite{circ}, given by:
\begin{equation}
F(x) = - \alpha x + \beta \ g(x)
\end{equation}
with the piecewise-linear function $g (x)=x$         when $x^*_{l} \le x \le x^*_{u}$, 
$g(x) = x^*_{l}$  when $x < x^*_{l}$ and 
$g(x)= x^*_{u}$   when $x > x^*_{u}$, 
where $x^*_{u}$ and $x^*_{l}$ are the upper and lower thresholds respectively.

We simulated the coupled dynamics of these two bi-stable systems for different network topologies as well. We find that the qualitative trends in both these bi-stable systems is similar to that described above, indicating the generality of the central results presented here.\\

{\bf Conclusions:}\\

        In summary, we have investigated the collective dynamics of bi-stable elements connected in different network topologies, ranging from rings and small-world networks, to scale-free networks and stars. We estimated the dynamical robustness of such networks by introducing a variant of the concept of multi-node basin stability which allowed us to gauge the global stability of the dynamics of the network in response to local perturbations affecting particular nodes of a system. We show that perturbing nodes with high closeness and betweeness-centrality significantly reduces the capacity of the system to return to the desired stable state. This effect is very pronounced for a star network which has one hub node with significantly different closeness/betweeness-centrality than the peripheral nodes. Considering such a network with all nodes in one well, if one perturbs the hub to another well, this {\em single} perturbed node drags the entire system to its well, thereby preventing the network from recovering its dynamical state. In contrast, even when {\em all} peripheral nodes are kicked to the other well, the hub manages to restore the entire system back to the original well. Lastly we explore explore Random Scale-Free Networks of bi-stable dynamical elements. Since the distribution of betweeness centralities, closeness centralities and degrees of the nodes is significantly different for Random Scale-Free Networks with $m=1$ and $m=2$, these networks have the potential to provide a test-bed for determining which of these centrality properties most inluences the robustness of the collective dynamics. The comparison between the global stability of these two classes of networks provides clear indications that the {\em betweeness centrality of the perturbed node is more crucial for dynamical robustness, than closeness centrality or degree of the node}. This result is important in deciding which nodes to safeguard in order to maintain the collective state of this network against targetted localized attacks. 


\end{document}